\documentclass[a4paper,11pt,bibtotocnumbered,english]{scrartcl}
\usepackage[a4paper, left=3cm, right=2.5cm, top=2.5cm, bottom=4cm]{geometry}
\usepackage{microtype}
\usepackage[ngerman,main=english]{babel}
\usepackage[utf8]{inputenc} %Codierung des LaTeX-Dokumentes. Auf Windows-Maschinen ist statt utf8 auch ANSIC als Codierung möglich, aber unnötig, da utf8 in jeder Hinsicht besser als ANSI ist. Bei Linux: latin1 als Codierung, auf MacOS X: applemac
\usepackage[T1]{fontenc}
\usepackage{amsmath, amssymb,amsfonts} %AMS-TeX-Pakete. Nötig für die Definition der Mathematik-Umgebung
\usepackage{graphicx} %Nötig, um Grafiken einbinden zu können
\usepackage[bookmarks,colorlinks=true]{hyperref} %Mittels hyperref lassen sich hyperlinks innerhalb des PDF-Dokumentes benutzen. Beispiel: Mausklick im Inhaltsverzeichnis auf ein Kapitel führt zum automatischen Sprung in dieses Kapitel
\usepackage{geometry}
\usepackage{float}
\usepackage{pdfpages}
\usepackage{framed, color} %Framed: Paket, mittels dessen ein Rahmen um einen Bereich definiert werden kann. Color: ä¤sst Farbdarstellung in Schrift, Hintergrund etc. zu
\usepackage{siunitx} %Ein schönes Paket, um Einheiten und physikalische Größen richtig zu setzen. Z.B.  \SI{2}{\kilo\gram\per\meter\squared}
\usepackage[square,numbers]{natbib}
\usepackage{gensymb}
\usepackage{hyperref}
\usepackage{caption}
\usepackage{fancyhdr}
\usepackage[none]{hyphenat}
\usepackage{svg}
\usepackage[ruled,linesnumbered]{algorithm2e}
\usepackage{subcaption}
\SetKw{KwBy}{by}
\SetKwComment{Comment}{// }{}
\usepackage{adjustbox}

\newcommand{\appendixpagenumbering}{
  \break
  \pagenumbering{arabic}
  \renewcommand{\thepage}{\thesection-\arabic{page}}
}

\DeclareSIUnit\curie{Ci}%Zusätzliche Einheit definieren.
\hypersetup{
	colorlinks,
	linktocpage,
	citecolor=black,
	filecolor=black,
	linkcolor=black,
	urlcolor=black
}
\graphicspath{{./graphics/}}

\counterwithin{figure}{section}
\counterwithin{table}{section}

\numberwithin{equation}{section} % Die Nummerierung von Gleichungen bekommt die jeweilige Section-Nummer als PrÃ¤fix

\newcommand{\Titel}{GPU Implementation of the Wavelet Tree}
\newcommand{\TitelDE}{GPU-Implementierung des Waveletbaums}

\begin{document}
\thispagestyle{empty}
\fancyhf{}
\renewcommand{\headrulewidth}{0pt}
\fancyfoot[R]{\thepage}
\fancyfoot[C]{\rightmark}

\begin{titlepage}

\begin{center}
\Large{\textbf{\Titel}}\\% \Huge \huge \Large \normalsize \Small usw. bestimmt die Schriftgröße.
\vspace{10mm}
\Large{\textbf{\TitelDE}}
\vspace{30mm}% Abstand

\large{Master's thesis}\\
of\\
Marco Franzreb Salgado\\
\vspace{60mm}
carried out at\\
Texas State University\\
under supervision of\\
Prof. Dr. Martin Burtscher\\
second supervisor and examiner\\
Priv.-Doz. Dr.-Ing. Stephan Rudolph\\
\vspace{30mm}
Stuttgart, 5th of May 2025\\
\end{center}

\end{titlepage}
% Hier die PDF der unterschriebenen Aufgabenstellung einfügen
% \includepdf[pages=-,angle=90]{Aufgabenstellung Unterschrieben.pdf}
\pagenumbering{Roman} \setcounter{page}{1}
\newpage\null\thispagestyle{empty}\newpage % Leerseite
\section*{Statement of Originality}
This thesis has been performed independently with support of my supervisor. It contains no material that has been accepted for the award of a degree at this or any other university. To the best of my knowledge and belief, this thesis contains no material previously published or written by another person except where due reference is made in the text. I further declare that I have performed this thesis according to the existing copyright policy and the rules of good scientific practice. In case this work contains contributions of someone else (eg. pictures, drawings, text passages etc.), I have clearly identified the source of these contributions, and, if necessary, have obtained approval from the originator for making use of them in this thesis. I am aware that I have to bear the consequences in case I have contravened these duties.

\includegraphics[scale=0.2]{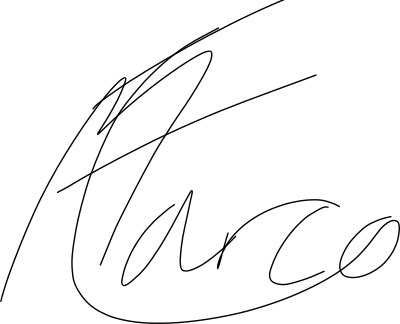}

05.05.2025

\subsection*{Erklärung}
\begin{otherlanguage}{ngerman}
Hiermit versichere ich, dass ich diese Masterarbeit selbstständig mit Unterstützung des Betreu\-ers angefertigt und keine anderen als die angegebenen Quellen und Hilfsmittel verwendet habe. Die Arbeit oder wesentliche Bestandteile davon sind weder an dieser noch an einer anderen Bildungseinrichtung bereits zur Erlangung eines Abschlusses eingereicht worden. Ich erkläre weiterhin, bei der Erstellung der Arbeit die einschlägigen Bestimmungen zum Urheberschutz fremder Beiträge entsprechend den Regeln guter wissenschaftlicher Praxis eingehalten zu haben. Soweit meine Arbeit fremde Beiträge (z.B. Bilder, Zeichnungen, Textpassagen etc.) enthält, habe ich diese Beiträge als solche gekennzeichnet (Zitat, Que\-llen\-an\-ga\-be) und eventuell erforderlich gewordene Zustimmungen der Urheber zur Nutzung dieser Beiträge in meiner Arbeit eingeholt. Mir ist bekannt, dass ich im Falle einer schuldhaften Verletzung dieser Pflichten die daraus entstehenden Konsequenzen zu tragen habe.
\end{otherlanguage}

\includegraphics[scale=0.2]{graphics/Signature.png}

05.05.2025

\addcontentsline{toc}{section}{\protect\numberline{}Statement of Originality}
\newpage\null\thispagestyle{empty}\newpage
\section*{Kurzfassung}
\addcontentsline{toc}{section}{\protect\numberline{}Kurzfassung}
\begin{otherlanguage}{ngerman}
Der Waveletbaum ist eine kompakte Datenstruktur zur Darstellung einer Folge von Werten. Der Begriff ``kompakt'' bedeutet, dass er nahezu die informationstheoretische untere Schranke hinsichtlich des Speicherplatzes erreicht. Er wird üblicherweise zur Darstellung von Text verwendet, kann aber auch eine Neuordnung oder ein Punktgitter repräsentieren~\cite{wt_for_all}. Der Vorteil der Verwendung eines Waveletbaums gegenüber der ursprünglichen Wertefolge liegt darin, dass er zwei Arten von Anfragen sehr effizient unterstützt, während er dennoch den Zugriff auf beliebige Werte der Folge erlaubt. Für eine gegebene Folge $S$ mit $n$ Werten unterstützt der Waveletbaum folgende Anfragen:

\begin{itemize} 
\item \textbf{$access(i)$}: Für $i \in [0, n)$, gibt den Wert an Position $i$ zurück.

\item \textbf{$rank_v(i)$}: Für $i \in [0, n]$ und einen Wert $v$, gibt die Anzahl der Vorkommen von $v$ in $S[0, i)$ zurück.

\item \textbf{$select_v(i)$}: Für $i \geq 1$ und einen Wert $v$, gibt die Position des i-ten Vorkommens von $v$ zurück.
\end{itemize}

Aufgrund der Struktur des Baumes, auf die ich gleich eingehen werde, benötigen diese Anfragen nur $\mathcal{O}(lg \sigma)$ Zeit, wobei $lgx \equiv log_2(x)$ und $\sigma$ die Alphabetgröße ist, also die Anzahl der eindeutigen Werte, aus denen die Folge besteht. Beim Zugriff auf die ursprüngliche Wertefolge wäre der Zugriff eine Operation in konstanter Zeit. In diesem Fall ist die Verwendung des Waveletbaums also ineffizienter, dies ist jedoch akzeptabel, da darin nicht der Nutzen dieser Datenstruktur liegt. Die rank- und select-Anfragen hingegen würden $\mathcal{O}(n)$ Zeit benötigen, was den Waveletbaum bei der Verarbeitung dieser Anfragen deutlich effizienter macht.

Der Aufbau des Waveletbaums funktioniert wie folgt: Gegeben eine Folge $S$ der Länge $n$, bestehend aus Werten eines Alphabets $\Sigma$ der Größe $\sigma$, teilt der Waveletbaum das Alphabet rekursiv in zwei Hälften und weist den Werten von $S$, die vom linken Kindknoten abgedeckt werden, die Zahl $0$ zu, und $1$ andernfalls. Dies liegt daran, dass er als Binärbaum dargestellt wird – dies ist jedoch keine zwingende Einschränkung, er kann auch als Baum mit höherer Dimensionalität dargestellt werden, wie von Ceregini et al.~\cite{quadWT} beschrieben. Dies bedeutet, dass der Baum $\lceil lg\sigma\rceil$ Ebenen hat, was die Zeitkomplexität von $\mathcal{O}(lg\sigma)$ erklärt. Jede Ebene enthält ein Bit-Array, das höchstens $n$ Bits lang ist. Abbildung \ref{fig:pointer_wt} zeigt einen Waveletbaum eines Beispieltexts. Damit die Zeitkomplexität von $\mathcal{O}(lg\sigma)$ tatsächlich erreicht werden kann, muss jedoch jede Operation auf einer Ebene in konstanter Zeit ausführbar sein. Da auf jeder Ebene binäre rank- oder select-Anfragen durchgeführt werden, benötigt der Waveletbaum eine zusätzliche Datenstruktur zur Unterstützung von rank und select, damit diese in konstanter Zeit abgefragt werden können.

Eine der relevantesten Anwendungen des Waveletbaums ist als Hilfsstruktur im FM-Index~\cite{fm-index}, einem komprimierten Volltextindex. Dieser Index wird häufig in der DNA- und Protein- Sequenzalignment~\cite{Bowtie} verwendet – eine Anwendung, die maximalen Durchsatz erfordert, da Sequenziermaschinen extrem schnell sind und die Ausrichtung der Sequenzen der Engpass im Prozess ist~\cite{sequencing}. Für solche Anwendungen sind GPUs, durch ihren massiven Parallelismus, speziell gut geeignet.

Meines Wissens nach existiert derzeit nur eine öffentlich verfügbare GPU Implementierung des Waveletbaums, und zwar in der NVBIO-Bibliothek\footnote{\url{https://github.com/NVlabs/nvbio}}. Diese Bibliothek wurde jedoch vor 11 Jahren entwickelt und seit 5 Jahren nicht mehr gepflegt. Zudem habe ich es nicht geschafft, sie mit aktuellen Versionen des CUDA-Compilers zu kompilieren. Deswegen ist das Ziel der Masterarbeit eine effiziente GPU Implementierung dieser Datenstruktur zu schaffen, um Anwendungen die diesen Parallelismus ausnutzen können, den Einsatz zu erleichtern.

Die Implementierung der rank and select Datenstrukturen, die die Basis für den Waveletbaum sind, schafft eine 40- bis 130-fache Verbesserung in Konstruktionszeit gegenüber der CPU Implementierung (\textit{wide-popcount} aus~\cite{Kurpicz_RS}). Es ist allerdings wichtig zu beachten, dass der CPU-Konstruktionsalgorithmus sequentiell ist, während mein eigener von der hohen Pa\-ra\-lle\-li\-tät profitiert, die eine GPU bietet. Der Durchsatz von rank-Anfragen gegenüber der CPU Implementierung, welche die effizienteste für diese Anfragen ist, ist mindestens 10 mal höher. Wenn man select-Anfragen vergleicht, wofür, laut den Authoren, die \textit{wide-popcount} Version nicht geeignet ist, schafft meine Implementierung einen Durchsatz der mindestens 100 mal höher ist, und in manchen Fällen bis zu 3'700 mal höher. Verglichen mit der effizientesten Version aus~\cite{Kurpicz_RS} (\textit{flat-popcount}) für diese Anfragen, ist der Durchsatz meiner Implementierung mindestens 10 mal höher.

Verglichen mit den besten parallelen CPU Konstruktionsalgorithmen, ist mein Algorithmus um den Baum zu konstruieren etwas langsamer, da die zusätzliche Zeit um den Text von der CPU auf die GPU zu kopieren einen Großteil der Laufzeit ausmacht. Wenn man die Zeit um den Text zu kopieren nicht berücksichtigt, ist mein Algorithmus zwischen 2x und 5x schneller.

Für die access, rank und select queries, messe ich die Zeit die nötig ist um eine Anzahl an Anfragen zu bearbeiten inklusive die Zeit um diese Anfragen auf die GPU zu kopieren und die Ergebnisse wieder auf die CPU zurück zu kopieren. Den Vergleich mache ich gegenüber der Implementierung aus der Succinct Data Structures Library (SDSL)\cite{SDSL}, welche die meist benutze ist. Für alle getesteten Texte ist meine Implementierung ab 500'000 Anfragen schneller. Für Anzahlen an Anfragen die den vollen parallelismus ausnutzen, ist meine Implementierung über 4 mal schneller.
\end{otherlanguage}

\newpage

\section*{Abstract}
I present a new GPU implementation of the wavelet tree data structure. It includes binary rank and select support structures that provide at least 10 times higher throughput of binary rank and select queries than the best publicly available CPU implementations at comparable storage overhead. My work also presents a new parallel tree construction algorithm that, when excluding the time to copy the data from the CPU to the GPU, outperforms the current state of the art. The GPU implementation, given enough parallelism, processes access, rank, and select queries at least 2x faster than the wavelet tree implementation contained in the widely used Succinct Data Structure Library (SDSL), including the time necessary to copy the queries from the CPU to the GPU and the results back to the CPU from the GPU.
\addcontentsline{toc}{section}{\protect\numberline{}Abstract}

\newpage\null\thispagestyle{empty}\newpage
\tableofcontents
\addcontentsline{toc}{section}{\protect\numberline{}Contents}
\newpage

\section*{Abbreviations}
\addcontentsline{toc}{section}{\protect\numberline{}Abbreviations}
\begin{tabular}{l l}
\textbf{DNA} & Deoxyribonucleic Acid\\
\textbf{CPU} & Central Processing Unit\\
\textbf{GPU} & Graphics Processing Unit\\
\textbf{NVBIO} & NVIDIA Bioinformatics Library\\
\textbf{CUDA} & Compute Unified Device Architecture\\
\textbf{SDSL} & Succinct Data Structures Library\\
\textbf{CUB} & CUDA UnBound\\
\textbf{MSB} & Most Significant Bit\\
\textbf{LSB} & Least Significant Bit\\
\textbf{SM} & Streaming Multiprocessor\\
\textbf{BA} & Bit Array\\
\textbf{TPQ} & Threads Per Query\\
\textbf{GB} & Gigabyte\\
\textbf{API} & Application Programming Interface\\

\end{tabular}
\newpage
\listoffigures
\addcontentsline{toc}{section}{\protect\numberline{}List of Figures}
\newpage
\listoftables
\addcontentsline{toc}{section}{\protect\numberline{}List of Tables}
\newpage\null\thispagestyle{empty}\newpage
\pagenumbering{arabic} \setcounter{page}{1}
\section{Introduction}\label{intro}
The wavelet tree is a succinct data structure used for representing a sequence of values. The term succinct means that it uses space close to the information-theoretic lower bound. It is usually employed for representing text, but can be used as a reordering or point-grid~\cite{wt_for_all}. The advantage of using a wavelet tree instead of the original sequence of values is that it efficiently supports two queries while still allowing access to any value of the sequence. Given a sequence $S$ with $n$ values, these queries are:

\begin{itemize}
    \item \textbf{$access(i)$}: Given $i \in [0, n)$, returns the value at position $i$.
    
    \item \textbf{$rank_v(i)$}: Given $i \in [0, n]$ and a value $v$, returns the number of occurrences of $v$ in $S[0, i)$.

    \item \textbf{$select_v(i)$}: Given $i \geq 1$ and a value $v$, returns the position of the $i^{th}$ occurrence of $v$.
\end{itemize}

Because of how the tree is built, which I will go into shortly, these queries require $\mathcal{O}(lg \sigma)$ time, where $lgx \equiv log_2(x)$ and $\sigma$ is the alphabet size, which refers to the number of unique values the sequence is composed of. On the original sequence, accessing a value would be a constant-time operation, so in this case using the wavelet tree is less efficient, but this is acceptable since it is not where the usefulness of the data structure lies. The rank and select queries, however, would require $\mathcal{O}(n)$ time, making the wavelet tree significantly more efficient at processing these queries. I go into more detail of how these queries work in Sections \ref{access_sec}, \ref{rank_sec}, and \ref{select_sec} respectively.

The structure of the wavelet tree works as follows. Given a sequence $S$ of length $n$ composed of values from an alphabet $\Sigma$ with size $\sigma$, the wavelet tree recursively splits the alphabet into two and assigns $0$ to the values of $S$ that are covered by the left child node, and $1$ otherwise. This is because it is represented as a binary tree, but there is no restriction to that and it could be represented as a multi-ary tree as well, as described by Ceregini et al.~\cite{quadWT}. This means that the tree will have $\lceil lg(\sigma)\rceil$ levels, which is where the $\mathcal{O}(lg\sigma)$ time complexity comes from. Each level then contains an array of bits that is at most $n$ bits long. Figure \ref{fig:pointer_wt} shows a wavelet tree of an example text. However, the $\mathcal{O}(lg\sigma)$ time complexity requires that whatever is happening at every level be of constant time complexity. Since we are performing binary rank or select queries at every level, the wavelet tree also requires a rank and select support structure in order to make these queries be of constant time complexity. These are described in more detail in Section \ref{RS_struct}. The construction of the tree is covered in depth in Section \ref{WT_sec}.

Since the goal of this thesis is to implement the wavelet tree for usage in GPUs, I must first answer the question of why this is even necessary. GPUs are parallel high-throughput machines, so the only use cases that would really benefit from them are the ones that perform many queries in parallel. One of the currently most relevant uses of the wavelet tree is as a support structure of the FM-index~\cite{fm-index}, which is a compressed full-text index. This index is commonly used in DNA and protein sequence alignment~\cite{Bowtie}, which is an application that requires maximal throughput, since the sequencing machines are extremely fast and the alignment of the sequences is the bottleneck of the process~\cite{sequencing}. Currently, to my knowledge, there is only one publicly available GPU implementation of the wavelet tree, which is from the NVBIO library\footnote{\url{https://github.com/NVlabs/nvbio}}. This library, however, was implemented 11 years ago and has not been maintained for the past 5 years. I also did not manage to compile it on modern versions of the CUDA compiler. For this reason, I cannot include it in my comparisons.

The source code of my implementation is available in a GitHub repository\footnote{\url{https://github.com/mfranzreb/ECL\_wavelet}}.

\newpage
\section{Related work}\label{rel_work}
Grossi et al.~first introduced the wavelet tree in 2003~\cite{wt_og}. Since then, a lot of research has been done in order to find applications for the data structure, as well as to improve the construction and query times.

In~\cite{wt_for_all}, Navarro gives a clear and through explanation of the data structure, its variations, and how the queries on it are performed. He then describes multiple compression schemes that can be applied to it by encoding the bit arrays and/or changing the tree shape. Lastly, he describes the multiple use cases of the tree as a representation of sequences, reorderings, and point-grids. Makris~\cite{Makris2012-bi} also presents a survey of the research done until the point of it's writing, in 2012.

Since its inception, multiple algorithms have been devised for the construction of wavelet trees, and, more recently, the focus has shifted towards parallel construction algorithms. The current state of the art is presented by Dinklage et al. \cite{practical_WT}. In their publication, the authors describe multiple sequential and parallel algorithms in detail, with different performance and additional space requirements. They also reference many previous sequential and parallel algorithms, and use them in their experimental results as comparison. The underlying concepts used in their parallel algorithms are what inspired mine. 

There are also variations of the original wavelet tree data structure, like the wavelet matrix and Huffman versions of the tree and matrix, which are presented by Navarro~\cite{wavelet_matrix}. I will not describe them here, as they are not relevant to the current work, but the publication also contains a through description of the different variants of wavelet trees, namely the pointer-based and the level-wise variants. He also presents the algorithms for computing the queries depending on the variant used, which are the ones that I used when implementing the queries.

Since the main computation performed in the queries is in the binary rank and select queries, being able to perform them efficiently is paramount. For this reason, the bit arrays are extended with rank and select support structures. Kurpicz~\cite{Kurpicz_RS} presents the most efficient low-overhead rank \textbf{and} select support structures. He first describes the different ways to speed up the queries, and then thoroughly explains his implementation. He also performs a thorough comparison with many previous implementations. My implementation is based on the \textit{wide-popcount} one described in the publication.

In the area of compression, Ferragina et al.~\cite{WT_compr} give a complete and highly technical theoretical analysis of different compression algorithms based on wavelet trees.
\newpage

\newpage\null\thispagestyle{empty}\newpage
\section{Preliminaries}
Since this thesis describes a GPU implementation of the wavelet tree, it is important that one is familiar with GPU architecture and specific terminology, such as warps, thread-blocks, shared memory, and coalescing, for example. If these terms are unfamiliar to the reader, I highly recommend the first two chapters of this training series\footnote{\url{https://www.olcf.ornl.gov/cuda-training-series/}}.

During the optimization process and in the performance evaluations, I will mostly refer to differences in percentages. For clarity, I will give a few examples now of how these percentages are to be interpreted. If I refer to something being slower or indicating that its value is higher, then the percentages are intuitive. If $x$ is 100\% slower than $y$, then $x = 2y$, if it is 35\% slower, then $x = 1.35y$. If I refer to something being faster or indicating that its value is lower, it is different. If $x$ is 100\% faster than $y$, then $x = 0$, if it is 35\% faster, then $x = (1-0.35)y$.
\newpage

\newpage\null\thispagestyle{empty}\newpage
\section{The Rank and Select Structure}\label{RS_struct}
\begin{figure}[H]
    \centering
    \includesvg[width=.7\linewidth]{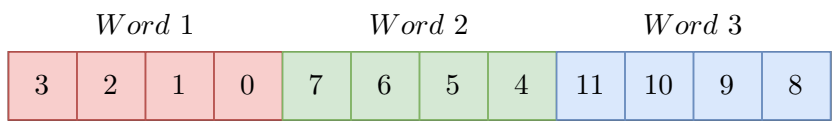}
    \caption{Example of bit storage inside words in the bit array, using 4-bit words.}
    \label{fig:BA_words}
\end{figure}

Before I introduce the rank and select support structure, I will first briefly describe the bit array and its implementation. Since the usual word size in a GPU is 32 bits and the registers have a size of one word, the array is implemented as an array of 32-bit unsigned integers. Inside each word, the bits are stored in reverse order, so that the first bit in the word is the least significant bit (LSB), which reduces the necessary operations when accessing a specific bit. Figure \ref{fig:BA_words} illustrates this using 4-bit long words.

In order to avoid having multiple bit array objects, and having to do multiple allocations, all the bit arrays of the wavelet tree are stored in one contiguous array. The starting indices of the bit arrays are then saved in another array. For increased performance, the start of each bit array is aligned to the beginning of a cache line, which is 128 bytes long.

The wavelet tree contains a bit array in each of its nodes, and for answering the queries, binary rank and/or select queries are performed on the bit arrays when traversing the tree. The \textit{rank} and \textit{select} of a bit array $B$ of length $n$ and $\alpha \in \{0, 1\}$, are defined as:

\begin{itemize}
    \item \textbf{rank}: Given $i \in [0, n]$, returns the number of ones (or zeros) in $B[0, i)$.

    \item \textbf{select}: Given a number $i$ starting from 1, returns the position of the i-th one (or zero).
\end{itemize}

During my research, I found that the binary rank was defined in different ways. Sometimes the upper bound was inclusive, and other times exclusive. The above definition follows the Succinct Data Structures Library's (SDSL)\cite{SDSL}, which contains the most widely used wavelet tree implementation.

When performing a wavelet tree query, the naive implementation would have an asymptotic time complexity of $\mathcal{O}(n)$ at each level, leading to worse asymptotic performance than performing the queries on the original text. The binary rank and select structures allow for constant-time binary queries, which improve the asymptotic-time complexity of a query from $\mathcal{O}(n\lceil\log\sigma\rceil)$ to $\mathcal{O}(\lceil\log\sigma\rceil)$, where $\sigma$ is the size of the alphabet. In the following sections, I will describe the implementation and performance improvements to the rank and select support structures.

\subsection{Binary Rank Support}
The binary rank support is based on the \textit{wide-popcount} implementation described in \cite{Kurpicz_RS}, since it is the best performing one for rank queries and well suited for GPUs. In this implementation, the bit array is split up into blocks of 65'536 bits, and the number of ones up to the beginning of each block is saved in a 64-bit unsigned integer. These blocks are called L1-blocks. If the number of zeros until the beginning of a block is needed, only a subtraction of the number of ones from the number of bits until the beginning of that block is necessary. Each L1-block is then split up into a certain number of smaller blocks, called L2-blocks, and the number of ones from the start of the L1-block up to the beginning of each L2-block is saved in a 16-bit unsigned integer.

The size of the L1-block was chosen for two reasons. First, it requires that the size of the integer for the L2-blocks be 16 bits, which is half a 32-bit word, and therefore simplifies access. Second, the size is a power of two, which greatly speeds up the division and modulo operations that are required to perform the binary rank queries.

The size of each L2-block can be modified to achieve the ideal performance-to-overhead ratio. I will dive deeper into this in Section \ref{access_sec}. The overhead of the support structure is then calculated as:

\[
\frac{\left\lceil \frac{n}{65'536} \right\rceil \times 64 + \left\lceil \frac{n}{|L2|} \right\rceil \times 16}{n}
\]

where $n$ is the size (number of bits) of the bit array and $|L2|$ the size of an L2 block. For reference, having $|L2| = 512$ results in an overhead of around 3.22\%.

Algorithm \ref{alg:bin_rank} illustrates how a binary $rank_1$ query is performed using the rank support structure. For a $rank_0$ query, only minimal changes are necessary. In the algorithm, edge cases are not considered. For example if $i=0$, then we know that the result will always be $0$. In Section \ref{access_sec}, I go through the optimization process.

\begin{algorithm}[hbt!]
\caption{Binary $rank_1$ using the rank support structure}\label{alg:bin_rank}
\KwIn{Bit array $B$ of length $n$ and index $i \in [0, n]$.}
\KwOut{Number of ones in [0, i).}
$result \gets 0$\;
$L1_i \gets \lfloor i / |L1| \rfloor$\ \Comment*{L1-block i is in.}
$result \gets result + L1[L1_i]$\;
$L2_i \gets \lfloor i / |L2|\rfloor$\ \Comment*{L2-block i is in.}
$result \gets result + L2[L2_i]$\;
$start\_bit \gets L2_i \times |L2|$\;
$end\_word \gets \lfloor i / 32\rfloor$\;
\For{$j\gets start\_bit$ \KwTo $i$ \KwBy $32$}{
    $word \gets word\_at\_bit(j)$\;
    \If{$\lfloor j/32\rfloor == end\_word$}
        {
        \Comment{Clear bits that are outside the range.}
        $word \gets get\_partial\_word(word, i\bmod 32)$\;
        }
    $result \gets result + popcount(word)$\;
}
\Return result\;
\end{algorithm}

\subsubsection{Rank support construction}
The nested partitioning of the bit array into L1- and L2-blocks maps nicely to the blocks and threads of a GPU. The number of ones in each L1-block can be computed in parallel, followed by an exclusive prefix sum. The same applies to the L2-blocks inside an L1-block. This means that each thread-block gets assigned an L1-block. Its work is then divided into two stages. First, we calculate the number of ones in each L2-block and save it into the corresponding array index. Then, we perform an exclusive prefix sum on the L2 array. The last thread of the prefix sum can then write the number of ones in the L1-block to the corresponding array position. After all thread blocks are done, an exclusive prefix sum yields the final result for the L1-blocks.

In order to speed up the computation there are multiple optimization options. I will first concentrate on the first part of the computation, where the number of ones in each L2-block gets computed. Here, we can add a temporary array to the shared memory of the thread-block for storing the number of ones in each L2-block, since it will be accessed multiple times. Additionally, instead of accessing the bit array word by word, we can perform the accesses two words at a time, which can increase memory throughput. 

Regarding the bit array accesses, there are two main possibilities. Either each thread in the block gets assigned an L2-block and then calculates the number of ones inside of it independently, or the number of threads per L2 block gets calculated such that there are as many threads as accesses necessary to compute the number of ones inside of the-L2 block. The first option (from now on option 1) does not take advantage of memory access coalescing, but does not need any communication between threads. The second option (option 2), on the other hand, takes full advantage of coalescing but requires a reduction so that the first thread in the group gets the total number of ones in the L2-block. When profiled on an NVIDIA A100, the results are surprising. Option 2 is 2\% slower, requiring 50\% more instructions due to the reduction, but having 30\% less warp stalling. Surprisingly, the memory throughput remains the same, and the L1 cache hit rate is 30\% less. Since it was slightly faster, I decided to stay with option 1.

For the second part, we need to compute the prefix sum of the L2 array and the total number of ones. I decided to perform the prefix sum, where each warp performs its own prefix sum over its slice of the array and saves it to shared memory. After all the chunks have been computed, it is only necessary to add the total number of ones in previous chunks to the entries in the current one. This can all be done with coalesced memory accesses and benefits from intra-warp communication when performing the prefix sum.

When computing the last L1-block, it is important to take into account that its size is arbitrary. The computation of the L1-blocks also produces as a result the total number of ones that are in the bit array, which will be useful for the construction of the select support structure.

\subsection{Binary Select Support}
A binary select query could be answered by finding the L1-block, and subsequent L2-block, that contain the sought after one (or zero). Using binary search, this is reasonably efficient but still dependent on the size of the bit array. A better solution is to save the position of every i-th one (and zero) as a 64-bit integer in an array. This constrains the possible L1-blocks in which the sought after one (or zero) can be. In this case, both ones and zeros have to be sampled, as the position of the i-th zero cannot be inferred from the sampled ones. How big the sampling interval should be can again be modified to achieve the ideal performance to overhead ratio. Since we are sampling both ones and zeros, the upper bound on the amount of samples that will be taken is the amount of one-samples in a bit array of the same size but full of ones. It can be calculated as follows:

\[
\frac{\left\lceil \frac{n}{i} \right\rceil \times 64}{n}
\]
where $n$ is the size (number of bits) of the bit array and $i$ is the sampling interval.

Algorithm \ref{alg:bin_select} illustrates how a binary $select_1$ query is performed using the rank and select support structures. For brevity, the algorithm shown does not take into account edge cases, like when there is only one L1-block or one L2-block. Figure \ref{alg:bin_select} illustrates how the algorithm works. At the top of the image, the samples are used to constrain the range of L1-blocks that have to be searched. This corresponds to lines two to 10. Once the L1-block has been found (line 11), the result is updated with the starting position of the L1-block (line 12), and the number of ones until its beginning is subtracted from the input index (line 13). Next, all the L2-blocks inside the L1-block are searched (lines 14-17). Once found, the result and input index are updated (lines 18 and 19). Finally, we count the number of ones inside the L2-block until the given index, giving us the final result (lines 20-31).

In Section \ref{select_sec}, I go through the optimization process.

\begin{algorithm}[p!]
\caption{Binary $select_1$ using the rank and select support structures}\label{alg:bin_select}
\KwIn{Bit array $B$ of length $n$ and i-th one $\in [1, n]$.}
\KwOut{Position of the i-th one.}
$result \gets 0$\;
\Comment{Index of nearest sampled one that is before i.}
$sample_i \gets \lfloor i / |sampling\_interval| \rfloor$\;
$result \gets result + samples[sample_i]$\;
$next\_sample\_pos \gets n$\;
\If{$(sample_i + 1) \times |sampling\_interval| <= |total\_num\_ones|$}{
    $next\_sample\_pos \gets samples[sample_i + 1]$\;
}
$total\_L1 \gets \lceil n / |L1|\rceil$\;

$start\_L1 \gets \lceil result / |L1| \rceil$\;
$end\_L1 \gets min(total\_L1, \lceil next\_sample\_pos / |L1| \rceil + 1)$\;
$wanted\_L1 \gets binary\_search(start\_L1, end\_L1)$\;
$result \gets wanted\_L1 \times |L1|$\;
\Comment{Convert to index starting from the L1 block.}
$i \gets i - L1[wanted\_L1]$\;

$total\_L2 \gets \lceil (n \pmod{|L1|}) / |L2|\rceil$\;
$start\_L2 \gets \lceil result / |L2| \rceil$\;
$end\_L2 \gets min(total\_L2, start\_L2 + |L1| / |L2|)$\;
$wanted\_L2 \gets binary\_search(start\_L2, end\_L2)$\;

$result \gets wanted\_L2 \times |L2|$\;
\Comment{Convert to index starting from the L2 block.}
$i \gets i - L2[wanted\_L2]$\;

$start\_bit \gets L2_i \times |L2|$\;
$num\_ones \gets 0$\;
\For{$j\gets start\_bit$ \KwTo $|L2|$ \KwBy $32$}{
    $word \gets word\_at\_bit(j)$\;
    \If{$num\_ones + popcount(word) \geq i \And num\_ones < i$}{
        $i \gets i - num\_ones$\;
        \Comment{Add the number of bits until the word and the position of the desired bit inside the word.}
        $result \gets result + j + getBitPos(word, i)$\;
        $break$\;
    }
    $num\_ones \gets num\_ones + popcount(word)$
}
\Return $result$\;
\end{algorithm}

\begin{figure}[H]
    \centering
    \includegraphics[width=.9\linewidth]{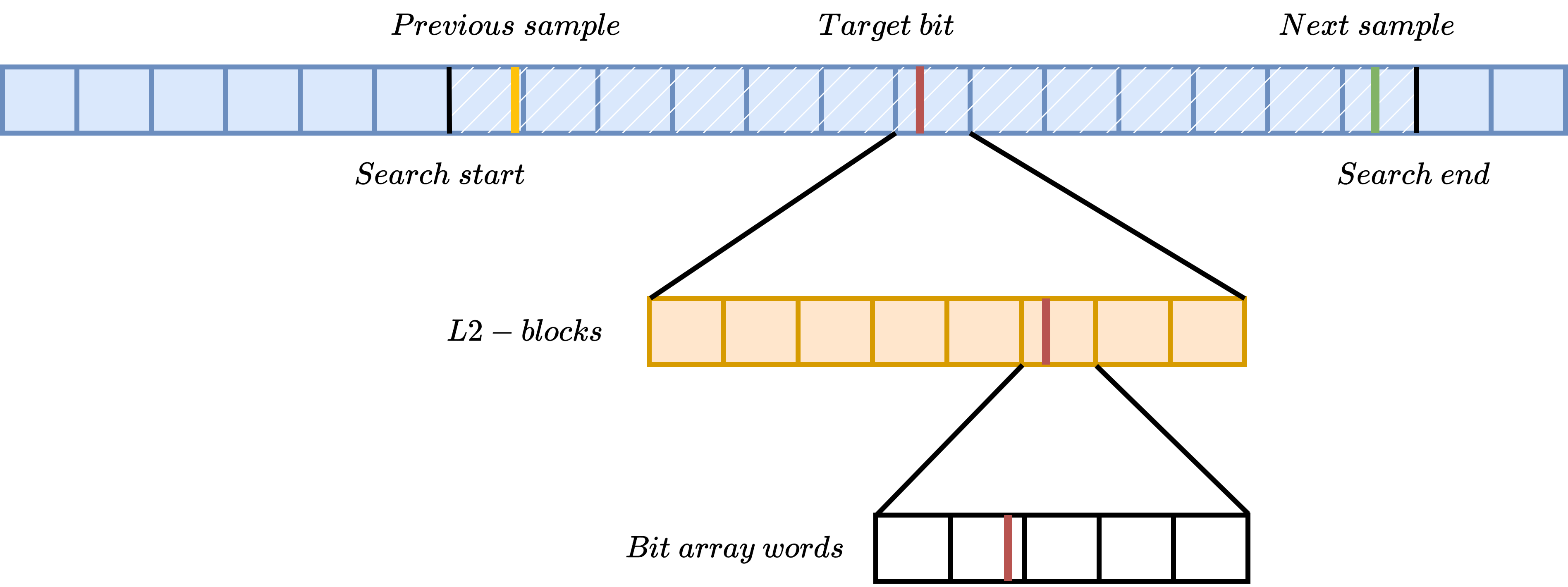}
    \caption{Illustration of the binary select algorithm using the rank and select structures.}
    \label{fig:bin_select}
\end{figure}

\subsubsection{Select support construction}
Once the total number of ones, and therefore zeros, has been computed by the rank support structure construction, the total number of one and zero samples is known. Constructing the select support structure is straightforward. Each thread takes a sample and computes the binary select of that sample to get its position, which it then writes into the array. Since the select support structure is being built, it cannot be used when computing the binary select of that sample. This is not a problem since the only difference is that the range of L1-blocks that needs to be searched is larger. The main work is performed in the binary select function, whose optimization is described in Section \ref{select_sec}.

\newpage

\newpage\null\thispagestyle{empty}\newpage
\section{The Wavelet Tree}\label{WT_sec}
\begin{figure}[H]
    \centering
    \begin{subfigure}{0.45\linewidth}
        \centering
        \adjustbox{valign=c}{\includegraphics[width=\linewidth]{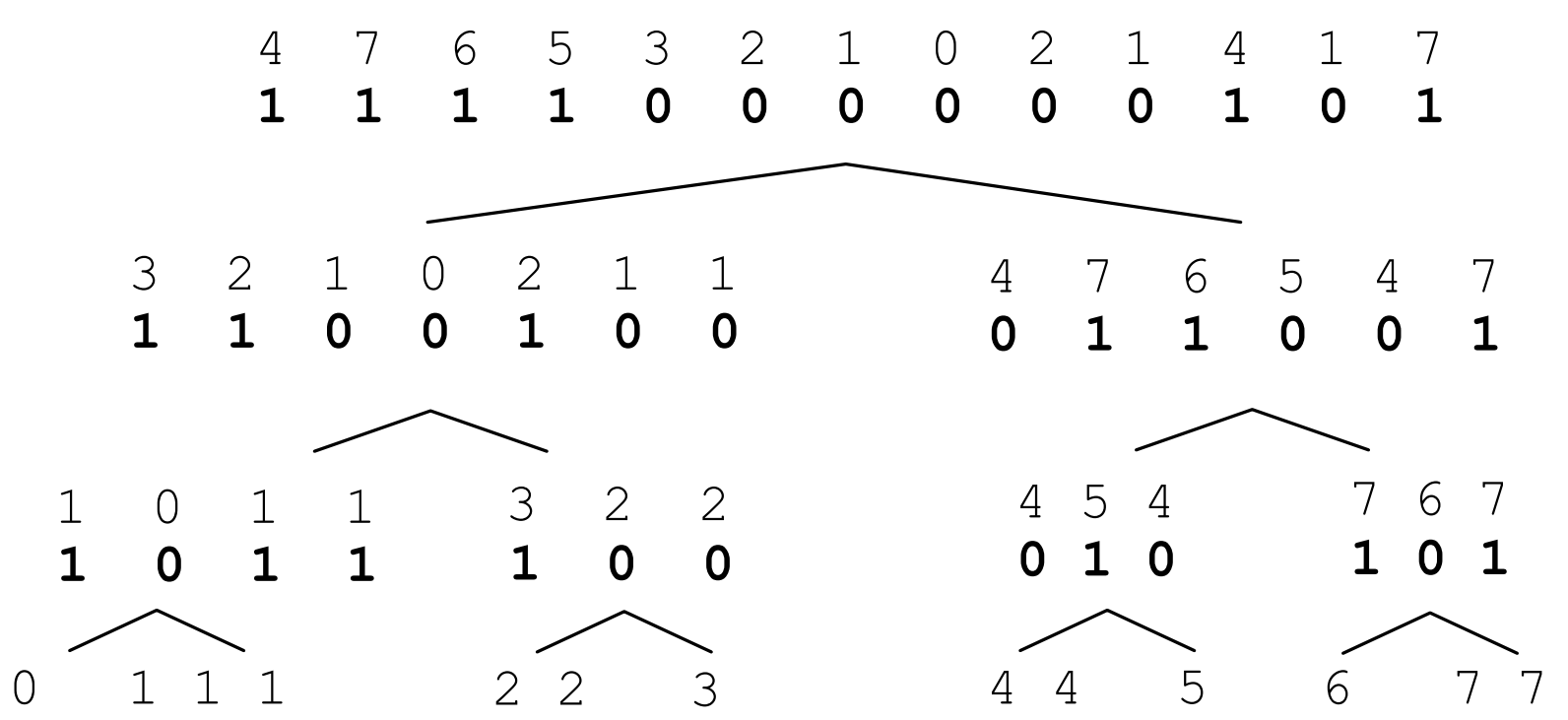}}
        \caption{Pointer-based tree}
        \label{fig:pointer_wt}
    \end{subfigure}
    \hfill
    \begin{subfigure}{0.45\linewidth}
        \centering
        \adjustbox{valign=c}{\includegraphics[width=\linewidth]{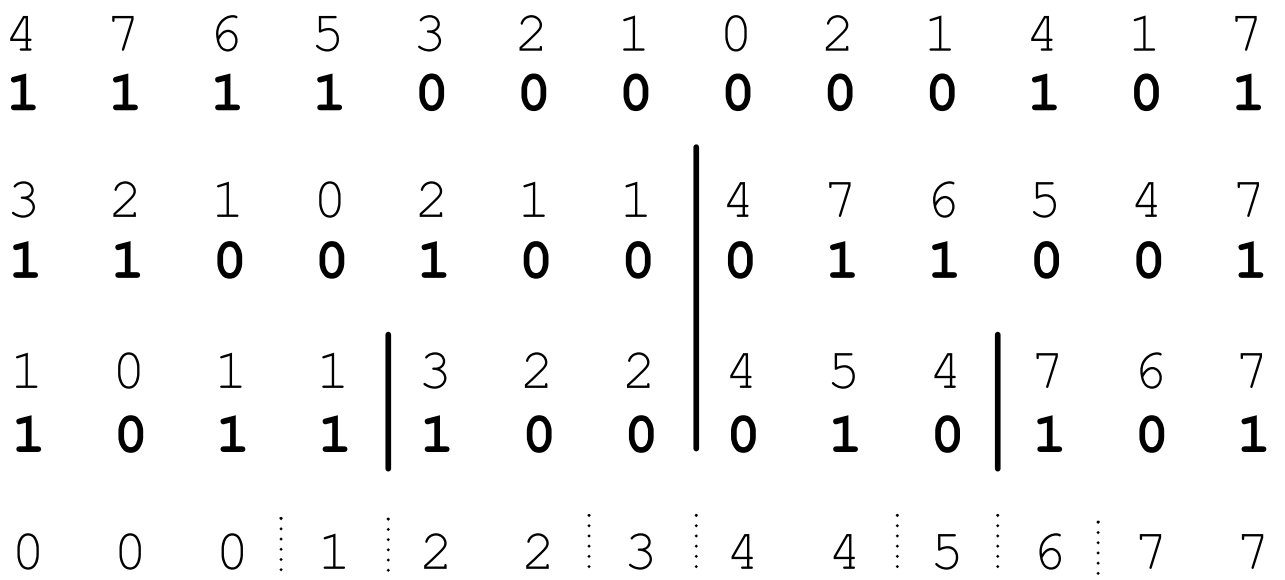}}
        \caption{Level-wise tree}
        \label{fig:level_wt}
    \end{subfigure}
    \caption{Figure taken from \cite{wavelet_matrix}. Only the bit arrays, in bold, and the tree topology are stored. For the level-wise extended version, we also store the cumulative histogram of the text, from which we can infer the node separations (bold vertical lines).}
    \label{fig:pointer_vs_level_wt}
\end{figure}

The wavelet tree has two main variants: the pointer-based \cite{Grossi_2003} and the level-wise \cite{MAKINEN2007332}. In the poin\-ter-ba\-sed variant, the nodes are separate and contain pointers to their children. In the level-wise variant, all nodes in a level are concatenated into one single bit array. This concatenation saves space and reduces redundancy in the rank and select structures that wrap the bit arrays. Figure \ref{fig:pointer_vs_level_wt}, taken from \cite{wavelet_matrix}, illustrates these variants. In order to improve the query operations of the level-wise wavelet tree, it can be extended by storing the exclusive prefix sum of the histogram of the text, which corresponds to the start of each node at the last level. With this information, the start of any node can also be inferred. Without this addition, the access and rank queries would require three binary ranks per level instead of one, and the select query two binary ranks per level instead of one. Additionally, for the select query, which would usually start at a leaf and traverse upwards, would have to start at the root of the tree in order to find the corresponding leaf and only then proceed back upwards. With the addition, the queries only require one extra binary rank per level, which can be pre-computed when constructing the tree, since we know that the position of these binary ranks will be at the start of the nodes. For these reasons, I chose to implement the extended variant of the level-wise wavelet tree. For more information on the other variants, please refer to \cite{wavelet_matrix}.

\subsection{Construction}
The algorithm for the construction of the tree is inspired by the parallel algorithms presented in \cite{practical_WT}. For their construction algorithms, the authors make use of a very important observation. If the alphabet of the wavelet tree is minimal, which means that the symbols are lexicographically ordered and inside the interval $[0, \sigma)$ where $\sigma$ is the alphabet size, the bits, starting from the most significant bit (MSB), decide the path the character takes at each level. You can see in Figure \ref{fig:pointer_wt} how this is the case. The fact that the alphabet has to be minimal is not a restriction since the text can be changed to a minimal alphabet efficiently.

This characteristic of the wavelet tree means that a level can be constructed by stably sorting the text according to the bits above the level. In case of the first level, since it doesn't have any bits above it, it just consists of the MSB of all the symbols in the text. For my implementation, I chose the radix sort, since it is a stable sorting algorithm that allows sorting by a specific subset of bits and is very efficient on GPUs. Moreover, the tree is computed from the top down, since it removes the necessity to keep the original text for the sorting at each level. This is due to the nature of the stable sort and minimal alphabet, since a character that, when sorted according to the $n$ MSBs comes before another, will never come after it when sorted according to the $n+1$ MSBs.

\begin{figure}[H]
    \centering
    \begin{subfigure}{0.45\linewidth}
        \centering
        \includegraphics[width=\linewidth] {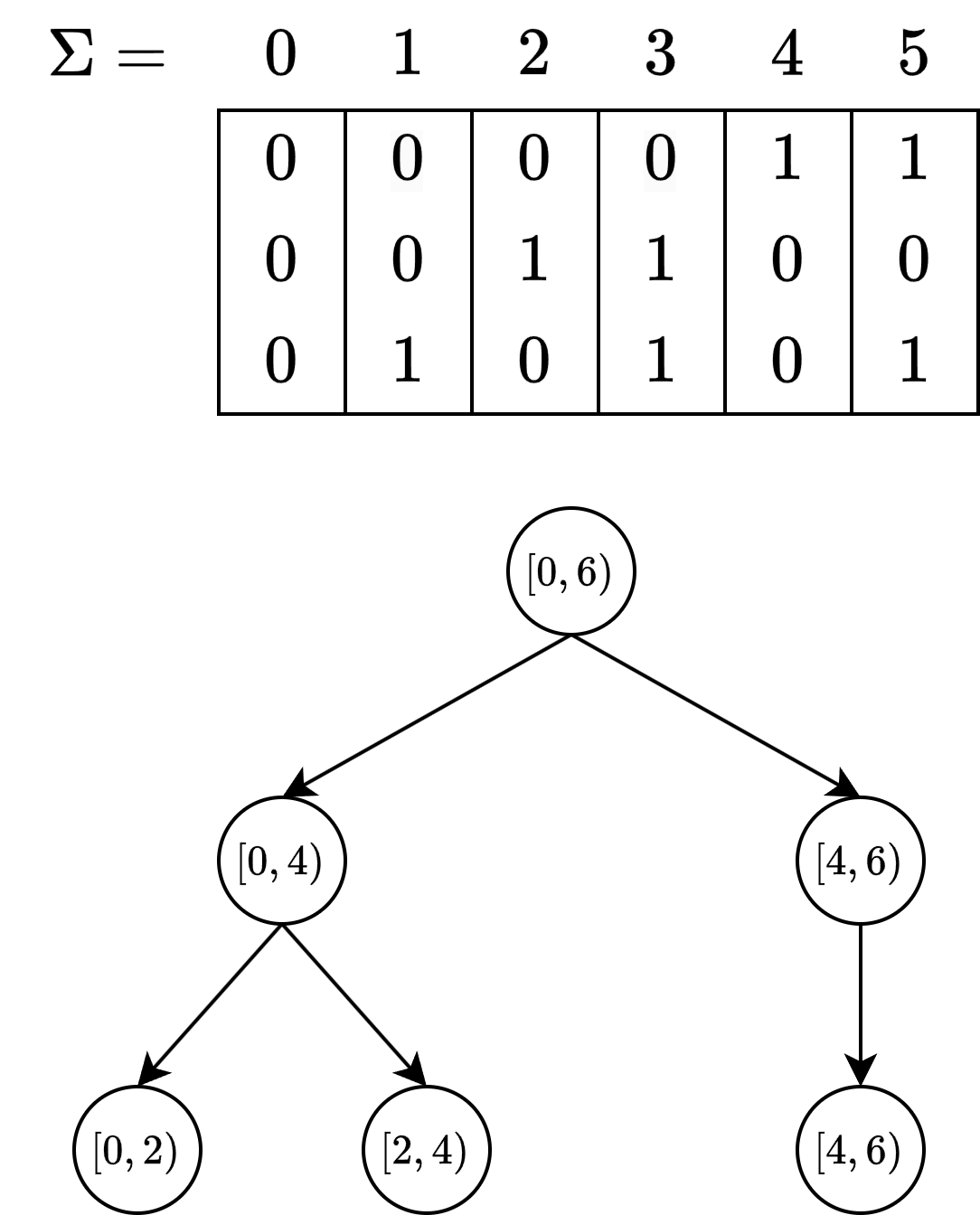}
        \caption{Redundant Wavelet Tree}
        \label{fig:redundant_wt}
    \end{subfigure}
    \hfill
    \begin{subfigure}{0.45\linewidth}
        \centering
        \includegraphics[width=\linewidth]{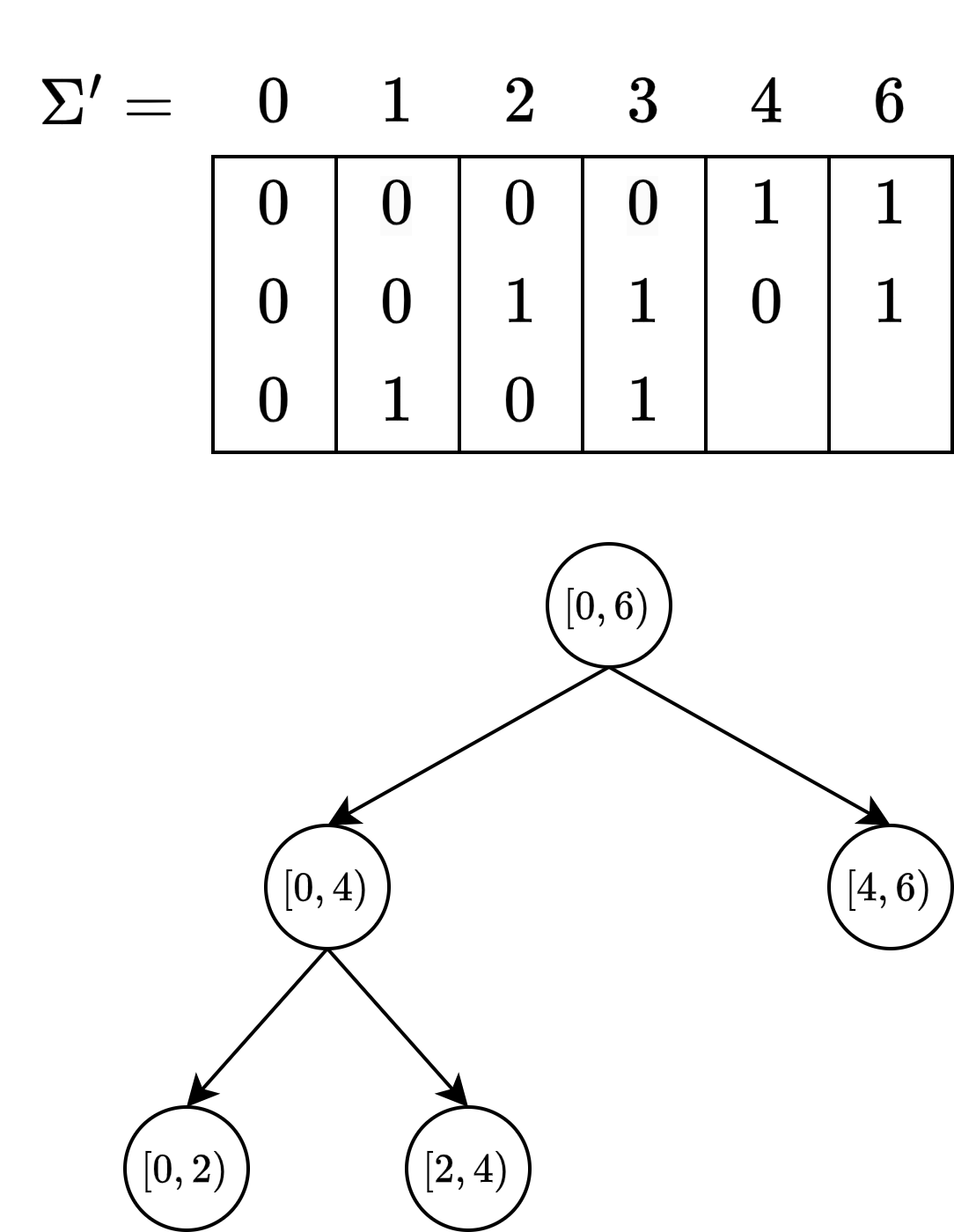}
        \caption{Reduced Wavelet Tree}
        \label{fig:reduced_wt}
    \end{subfigure}
    \caption{Comparison between the original tree structure when using the bits of the minimal alphabet ($\Sigma$) to create the structure, and the structure when codes are used for the symbols equal to or larger than the previous power of two of the alphabet size ($\Sigma'$). The intervals inside of the nodes denote the characters in the alphabet that are considered by the node. In both cases it refers to the minimal alphabet.}
    \label{fig:wavelet_tree_comparison}
\end{figure}

An undesirable effect from using the symbols' bits as the ones that dictate the structure of the tree is that, for alphabet sizes that are not a power of two, there will be nodes that have only one child instead of two. This makes the bit arrays at some of the levels larger than they should be, and increases traversal lengths in some cases. In order to combat this, the symbols that are bigger than the previous power of two of the alphabet size can be replaced by codes that ensure all nodes will have two children. This improvement results in a reduction of the size of the wavelet tree, as well as faster traversals in some cases, while only adding marginal complexity to the construction. Figure \ref{fig:wavelet_tree_comparison} compares the structure of the two trees for an alphabet size $\sigma = 6$. It is important to remember that the length of the codes needs to be saved along with the codes themselves, since they are different to the length of the original symbols.

When creating the codes, it is also important that their lengths be monotonically decreasing, since the nodes in a level are concatenated, which would not work if a leaf node was between two nodes. Algorithm \ref{alg:tree_codes} shows how the codes are created.

\begin{algorithm}[p!]
\caption{Create codes for minimal wavelet tree.}
\label{alg:tree_codes}
\KwIn{Integer $alphabet\_size$}
\KwOut{Vector of $codes$}
\BlankLine
\If{$isPowTwo(alphabet\_size)$}{
    \Return empty vector\;
}
$total\_num\_codes \gets alphabet\_size - prevPowTwo(alphabet\_size)$\;
Initialize $codes$ as a vector of size $total\_num\_codes$\;
$total\_num\_bits \gets \lceil \lg($alphabet\_size$) \rceil$\;
$alphabet\_start\_bit \gets total\_num\_bits - 1$\;
\BlankLine
\Comment{Start bit of the subset of codes we are looking at in an iteration.}
$start\_bit \gets 0$\;
\Comment{Start symbol of the subset of codes we are looking at in an iteration.}
$start\_i \gets 0$\;
$code\_len \gets total\_num\_bits$\;
$num\_codes \gets alphabet\_size$\;
\BlankLine
\Repeat{$code\_len == 1$}{
    \For{$i \gets code\_len - 1$ \KwTo $1$}{
        $pow\_two \gets 2^i$\;
        \If{$num\_codes \le pow\_two$}{
            \textbf{break}\;
        }
        $num\_codes \gets num\_codes - pow\_two$\;
        $start\_i \gets start\_i + pow\_two$\;
        $start\_bit \gets start\_bit + 1$\;
    }
    \BlankLine
    \If{$num\_codes == 1$}{
        $code\_len \gets 1$\;
        $codes[alphabet\_size-1].len \gets start\_bit$\;
        $codes[alphabet\_size-1].code \gets \Bigl((1 \ll start\_bit) - 1\Bigr) \ll \Bigl(alphabet\_start\_bit + 1 - start\_bit\Bigr)$\;
    }
    \Else{
        $code\_len \gets \lceil \lg(num\_codes) \rceil$\;
        \BlankLine
        \For{ $i \gets (alphabet\_size - num\_codes)$ \KwTo $alphabet\_size$}{
            $local\_code \gets i -  start\_i$\;
            $local\_code \gets local\_code \ll (total\_num\_bits - start\_bit - code\_len)$\;
            $local\_code \gets local\_code + (~((1 \ll (total\_num\_bits - start\_bit)) - 1))\ \&\ codes[i].code$\;
            $codes[i].code \gets local\_code$\;
            $codes[i].len \gets start\_bit + code\_len$\;
        }
    }
    \BlankLine
}

\BlankLine
\Return $codes$\;
\end{algorithm}

Going back to the construction of the tree, an extra step is necessary if codes are being used. Since the codes have different lengths than the other symbols of the alphabet, they need to be removed from the text whenever the level is reached where the code ends. In practice, the text does not need to be reduced, it is only necessary to reduce the portion of the text that should be input into the bit array at that level, since, due to the nature of the codes, all the codes that have reached their end will be at the end of the text. Algorithm \ref{alg:tree_construction} shows how the tree is constructed.

\begin{algorithm}[hbt!]
\caption{Wavelet tree construction.}
\label{alg:tree_construction}
\KwIn{Text $T$ of length $n$ with minimal alphabet of size $\sigma$.}
\KwOut{Wavelet tree of text.}
$codes \gets createCodes(\sigma)$\;
\Comment{Replace symbols with codes while computing the histogram, parallelized.}
$hist \gets computeHistogramAndEncodeText(codes, T)$\; 
$num\_levels \gets \lceil lg \sigma\rceil$\;
$level\_sizes \gets getLevelSizes(codes, hist)$\;
$cumulative\_hist \gets exclusivePrefixSum(hist)$\;

$fillLevel(T, n, 0)$\;
\For{$i\gets 1$ \KwTo $num\_levels$}{
    $radixSort(T, n, i)$\ \Comment*{Sort T using the bits $[MSB, i)$, parallelized.}
    \Comment{Fill the bit array at level i with the i-th MSB of each character in T, parallelized.}
    $fillLevel(T, level\_sizes[i], i)$\; 
}
$createRankSelectStructures()$\;
$precomputeRanks()$\;
\end{algorithm}

Apart from the space necessary for the tree and its supporting structures, the construction algorithm also needs space for the text, the sorted text, and additional space for the radix sorting algorithm that is dependent on the architecture of the GPU. For practicality, in case any of the additional memory that is not the tree itself does not fit fully on the GPU, it is allocated using unified memory. This allows the user to build a tree as long as the tree itself fits on the GPU, but using unified memory severely increases the time required to build the tree.

Now that the tree construction process has been covered, I will present the optimizations made to the most important sections of the construction. These are the histogram computation and the filling of a level. For the radix sort I used a highly optimized implementation by the CUB library \cite{CCCL}, so I will not mention it further.

\subsubsection{Histogram computation optimizations}
When computing the histogram of the text, I simultaneously replace the symbols with the minimal and encoded version of the alphabet if necessary. This means that I cannot use a library implementation, such as the one from CUB. Since replacing the symbols is a very straightforward operation, I will focus on the optimization of the histogram computation. 

The naive way to parallelize the computation is to have a grid-strided loop, where each thread accesses an adjacent element of the array and performs an atomic addition into the corresponding element of the histogram array. For big alphabets, this is not a bad option since, due to the large number of options, concurrent writes to the same element of the histogram will be minimal. However, when the alphabet is small, a large part of the writes will be serialized, negating most of the parallelism. A simple but efficient way of countering this involves using the GPU's shared memory. Using all of the available shared memory, as many block-local histograms as possible are initialized. The threads inside a thread-block then get assigned one of these shared memory histograms into which to write to. The assignment is done in a round-robin fashion in order to minimize the number of threads within a warp that share a local histogram. Even though the writes still need to be atomic, the contention is reduced significantly, and for very small alphabet sizes each thread has its own local histogram, eliminating any contention. If shared memory local histograms are used, they have to be reduced to the first local histogram, and then atomically added into the global histogram.

\begin{figure}[H]
    \centering
    \includegraphics[width=.9\linewidth]{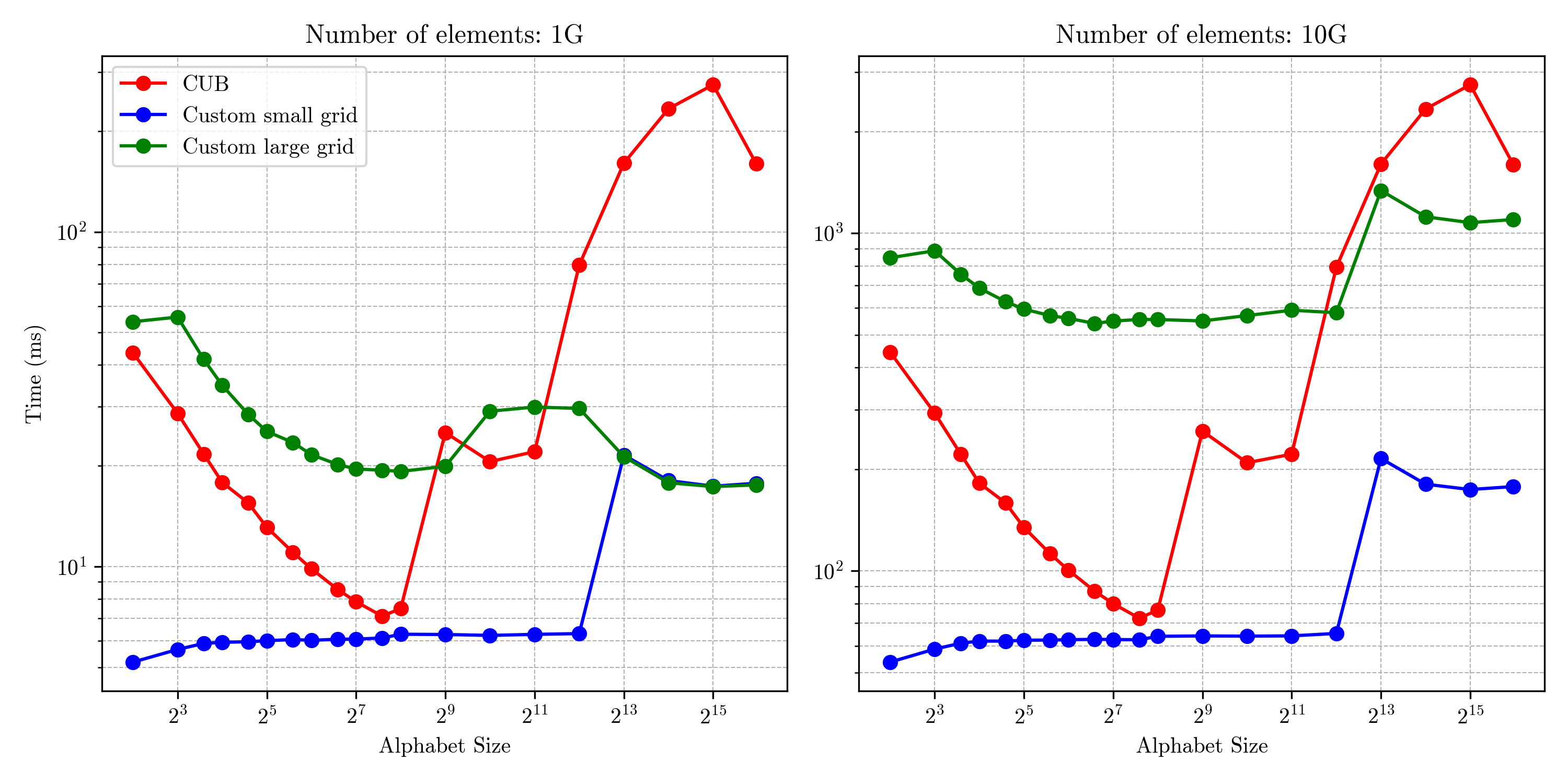}
    \caption{Comparison of CUB's histogram implementation with my own, which is either launched with as many threads as the size of the text (large grid) or as many threads as necessary to fully occupy the GPU (small grid). Texts are uniformly randomly generated. Run on an NVIDIA RTX 3090.}
    \label{fig:hist_comp}
\end{figure}

Another important optimization was finding the optimal total number of threads to launch. It is often recommended to launch as many threads as units of work (in this case the text size), but I found this to severely worsen performance in comparison to a persistent-thread approach, where I launch just enough threads in order to fully occupy the GPU. Figure \ref{fig:hist_comp} shows the difference between the different total numbers of threads for uniformly randomly distributed texts with a variety of alphabet sizes. This makes sense since the more threads you launch, and therefore blocks, the more atomic writes you need to make to the global histogram.

Surprisingly, this implementation performed significantly better than CUB's. Figure \ref{fig:hist_comp} also compares their performance for randomly distributed texts with a variety of alphabet sizes. The jump in execution time seen in my implementation corresponds to when the histogram becomes too large to fit in shared memory.

\subsubsection{Level filling optimizations}
When filling a level of the wavelet tree, it is paramount to maximize memory throughput, since it is the main bottleneck. For this, I found two possibilities.

In the first one, the threads access the array in a grid-strided loop and extract the desired bit from the character. A warp $\_\_ballot\_sync$ operation is then performed so that all threads inside a warp receive the word that needs to be written to the bit array, which is performed by the first thread in the warp. This results in a coalesced global read from the bit array, but an uncoalesced global write to it. An improvement on this uses shared memory so that the first thread in each warp writes its word to shared memory, and then a coalesced write to the bit array is performed. This results in a coalesced global read from the array, a bank-conflict free write and read to shared memory, and a coalesced global write.

\begin{figure}[H]
    \centering
    \includegraphics[width=.9\linewidth]{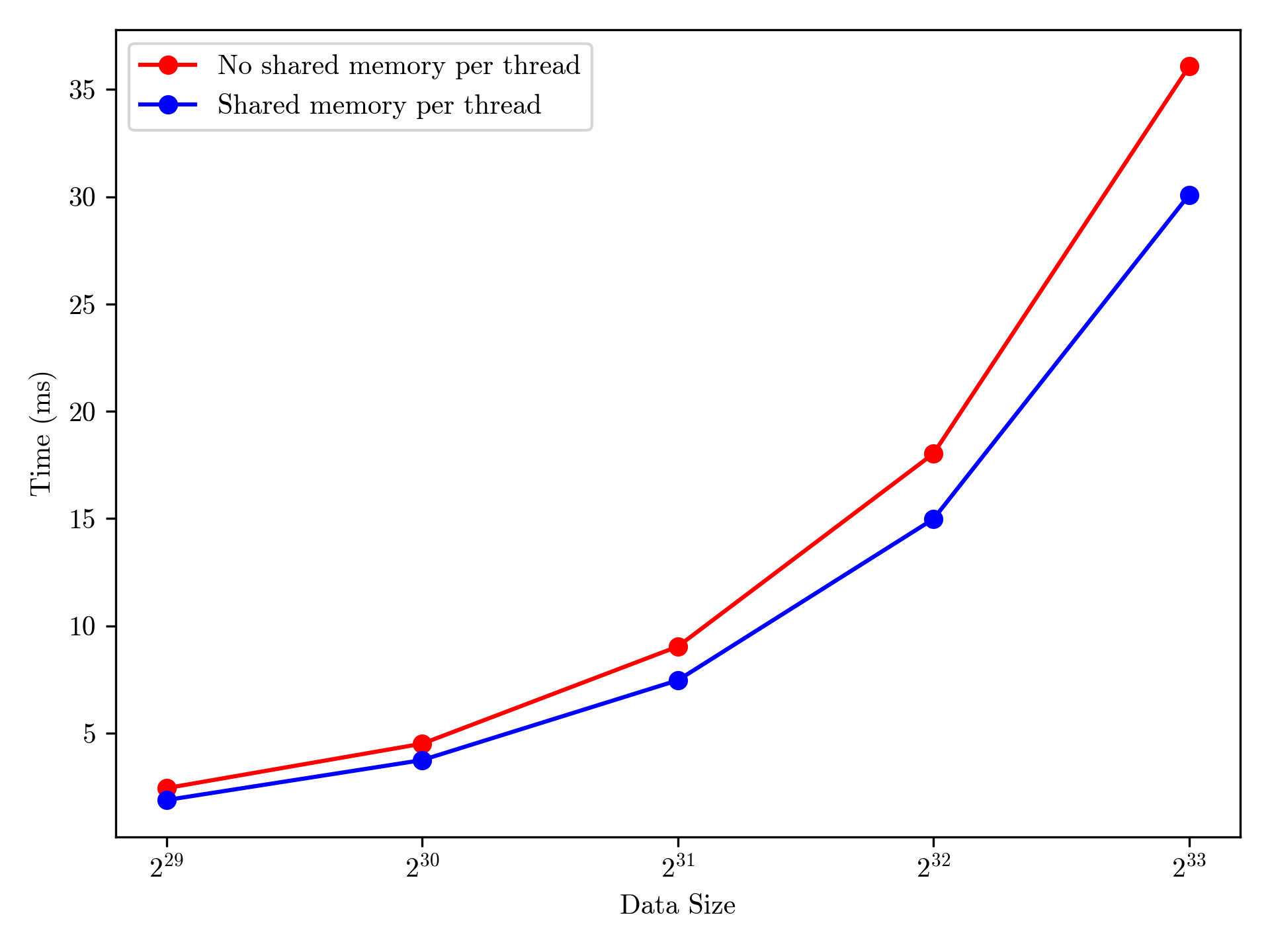}
    \caption{Comparison of the filling of a level of the wavelet tree using method one (no shared memory per thread) or method two (shared memory per thread). Texts are uniformly randomly generated with an alphabet size of 4. Run on an NVIDIA RTX 3090.}
    \label{fig:flk_comp}
\end{figure}

In the second possibility, each thread has an array in shared memory where it can store 32 symbols. The first step is to load a slice of the text into shared memory so that each thread has its 32 symbols in its shared memory array. Between arrays there is padding in order to avoid bank conflicts, since inside of a warp the threads will be accessing the same element in their array. Once each thread has its 32 symbols in shared memory, it accesses them sequentially and extracts the desired bit from each. It then writes the 32-bit word to the bit array. This results in fully coalesced accesses to global memory and bank-conflict free accesses to shared memory. The advantage of this method over the first is that the parallelism is at thread-level and not warp-level. Figure \ref{fig:flk_comp} shows a performance comparison of both methods.

In the implementation, if there is enough shared memory for the second option, it is then used. If not, the first option is used.

As with the histogram computation, the total number of threads launched had an impact on performance, so it was kept to the minimum necessary to achieve full occupancy.

\subsection{Access Query}\label{access_sec}
The algorithm for accessing the symbol at a specific index is shown in Algorithm \ref{alg:access}.

\begin{algorithm}[hbt!]
\caption{Wavelet tree access query.}
\label{alg:access}
\KwIn{Wavelet tree $wt$ and index $i$ of desired symbol.}
\KwOut{Symbol of text at index $i$.}
\Comment{Interval $[start\_symbol, end\_symbol)$ covered by current node.}
$start\_symbol \gets 0$\;
$end\_symbol \gets alphabet\_size$\;
\For{$l \gets 0$ \KwTo $wt.num\_levels$}{
\Comment{Number of occurrences of symbols smaller than $start\_symbol$. Corresponds to the start of the node in the level.} 
$counts \gets wt.cum\_histogram[start\_symbol]$\;
\If{$end\_symbol - start\_symbol \leq 2$}{
    \If{$end\_symbol - start\_symbol > 1\ \textbf{and}\ BA_l[counts+i] == 1$}{
        \Return $start\_symbol+1$\;
    } 
    \Else{
        \Return $start\_symbol$\;
    }
}
\Comment{Number of zeros until start of node.}
$start \gets wt.rank_0(l, counts)$ \;
\Comment{Number of zeros until desired index, $i$ is relative to the start of the node.}
$pos \gets wt.rank_0(l, counts + i)$ \;
\Comment{Width of the child node.}
$diff \gets prevPowTwo(end\_symbol - start\_symbol)$\;
\If{$BA_l[counts+i] == 0$}{
    $index \gets pos-start$\;
    $end\_symbol \gets start\_symbol + diff$\;
}
\Else{
    $index \gets index - (pos-start)$\;
    $start\_symbol \gets start\_symbol + diff$\;
}
}
\end{algorithm}

\begin{figure}[H]
    \centering
    \includesvg[width=.9\linewidth]{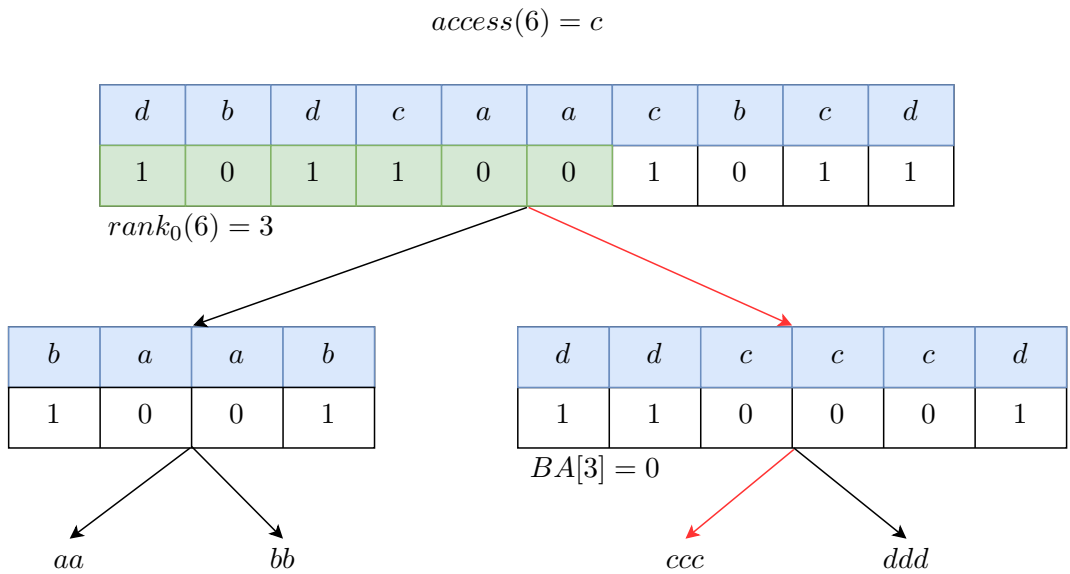}
    \caption{Example of accessing the symbol at index 6 using the wavelet tree of text $T = dbdcaacbcd$.}
    \label{fig:access_query}
\end{figure}

It is almost identical to the original algorithm (Algorithm 3 in \cite{wavelet_matrix}). The main improvement is that we do not need to traverse until the last level, since once there are only two possible symbols, we know which one is the desired one by accessing the bit that represents the symbol (lines 5-12). If the bit is one, then the desired symbol is the bigger one, otherwise, it is the smaller one. Figure \ref{fig:access_query} shows an example of an access query. The query itself already tells us which bit in the first level corresponds to the symbol we are looking for, we can use this information to know whether we need to traverse left or right. Before we can traverse to the child node, we need to find out at which position in that child node the desired character is located. For this, we count the number of zeros in the node before the position and subtract it from the position, which is equivalent to calculating the number of ones before that position. Having done that we now know that the bit corresponding to the desired character is at position 3 in the second node. Since now the only two options are $c$ or $d$, we only need to access the bit of that position to know if it is the left or right option. In this case, the bit at that position is zero, so we know that the desired symbol is $c$.

\subsubsection{Access query optimization}
In the implementation, there is also another optimization to the algorithm that is not included in Algorithm \ref{alg:access} for clarity. In line 16, the algorithm performs an access to the bit array, which can be spared since the binary rank operation in line 14 accesses that bit and can return it together with the result. When profiled, this change resulted in a 20-24\% improvement in runtime.

For processing a query, there are arrays that all threads in a block access at every level of the tree. This makes them good candidates for adding to shared memory. These are the offsets array, i.e., the array that contains information on where in the global array each level's bit array starts, and the cumulative histogram or counts array. Moving these arrays into shared memory improves the runtime by 5 to 10 \%. 

The binary rank computed in line 13 corresponds to the number of zeros until the start of a node. Since the tree will have no more than $\sigma$ nodes, where $\sigma$ is the alphabet size, these ranks can be pre-computed when constructing the tree, sparing one binary rank operation per level. If there is enough space available, the array can also be moved into shared memory. Surprisingly, when profiled on an NVIDIA RTX 2080 Ti, this resulted in a 5-10\% slowdown, despite 40\% fewer bit array accesses and 30\% fewer instructions executed. When profiled on an NVIDIA A100, however, this resulted in a runtime improvement of 5\% for $\sigma = 150$ and of 30\% for $\sigma = 20'000$. Since the improvement on the more recent GPU exceeded the slowdown on the older GPU, I decided to keep this optimization.

The binary ranks are the main work of the algorithm, so I will now describe their optimization process. Since the main work of the binary rank is performed in the loop (lines 8-14 in Algorithm \ref{alg:bin_rank}), that should be the focus of the optimization. 

The most impactful optimization was deciding how many threads should collaborate when processing a query, i.e., how many threads per query (TPQ) to have. Maximizing coalescing in the loop should result in ideal performance when processing a query. The downside of having multiple threads per query is that fewer queries can be processed in parallel. Throughout the optimization process, I tested each optimization with one, two, four, eight, 16 and 32 TPQ.

With the initial L2-block size of 2048 bits (64 words) where, on average, 32 loop iterations will be necessary, 32 TPQ should result in the best coalescing possible. However, when profiled, 16 TPQ was 16\% faster, due to a 25\% higher memory throughput and 33\% less instructions executed. The best configurations were four and two TPQ. When comparing two TPQ to 16 TPQ, it achieves 12\% higher memory throughput and 60\% less instructions executed. Interestingly, using one TPQ was almost twice slower than using two. This is mainly due to a 41\% lower L1 cache hit rate and only 14\% less instructions executed. The warp cycles per instruction also increase by 120\%, but throughout the optimization process I have found this metric to not be a good predictor of performance.

The simplest way to reduce time spent in the loop is to reduce the number of iterations, i.e., the number of words that have to be accessed. I therefore tested multiple L2-block sizes. When comparing a size of 2048 bits (64 words) with one of 1024 bits (32 words) for the best TPQ configurations (two and four), there is a 20\% improvement in runtime. The memory throughput increases by 9\%, but the L1 and L2 hit rate decrease by 12\%. The big difference is that now 34\% fewer instructions are executed. The one TPQ configuration improves by 34\% due to an increase in memory throughput of 13\% and 43\% fewer instructions executed. When going down to a size of 512 bits (16 words), compared to a size of 1024, the improvement is not as stark. The two and four TPQ configurations improve by 13\%, and the one TPQ configuration by 29\%, with the same trends in the performance metrics mentioned above. Understandably, the 32 TPQ configuration becomes slower, since half of the threads will not take part in the loop. When reducing the size to 256 bits (8 words), surprisingly, the two and four TPQ configurations change marginally, 2\% faster and 1\% slower respectively. The memory throughput barely changes, while the number of executed instructions is reduced by 17\%. On the other hand, the warp cycles per instruction increase also by 17\%. The one TPQ case, however, improves by 22\%. In order to keep the comparison to the CPU implementation fair, I kept the L2-block size at 512 bits.

The final main optimization to the implementation again focuses on reducing the number of iterations of the loop. In this case, each thread in the group accesses the bit array two words, or one 64-bit word, at a time. Interestingly, only the one TPQ configuration benefits from this change. It improves on average by 15\% and becomes the most performant configuration.

The next step in the optimization process was improving the launch configuration of the kernel. As with all other kernels mentioned, this one was also sensitive to the total number of threads launched. Launching as many threads as there are queries was slower than launching only as many threads as necessary to fully occupy the GPU. The minimum size of the thread-blocks is calculated so that it does not reduce the number of arrays that fit in shared memory. The exact size of the thread-blocks did not affect performance, as long as it remained a divisor of the maximum number of resident threads in a streaming multiprocessor (SM).

\begin{figure}[H]
    \centering
    \includesvg[width=.9\linewidth]{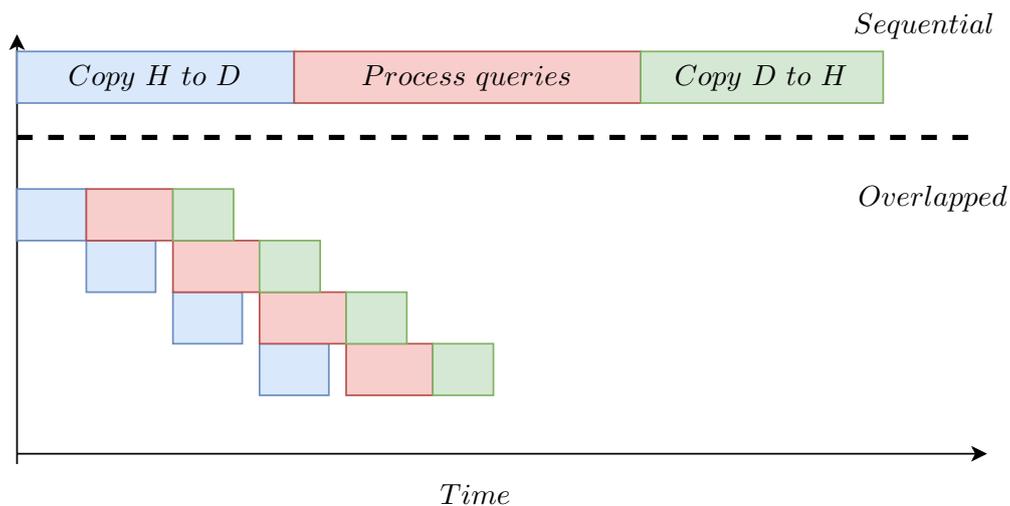}
    \caption{Illustration of the overlapping of copying and processing queries in order to minimize execution time. $H\ to\ D$ refers to the host-to-device copy of the queries, and $D\ to\ H$ to the copying of the results from the device back to the host.}
    \label{fig:query_launch}
\end{figure}

\begin{figure}[H]
    \centering
    \includesvg[width=.9\linewidth]{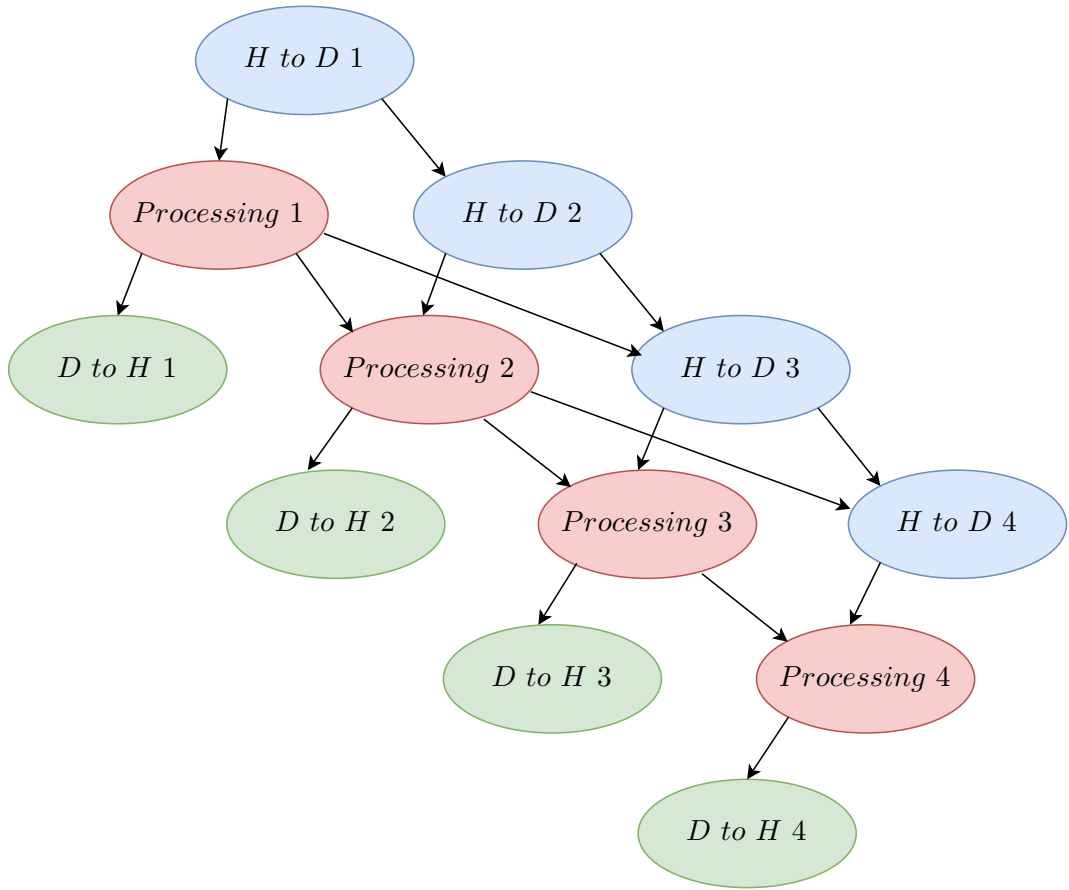}
    \caption{Graph of the launch ordering and dependencies for the processing of queries coming from the host. $H\ to\ D$ refers to the host-to-device copy of the queries, and $D\ to\ H$ to the copying of the results from the device back to the host.}
    \label{fig:query_graph}
\end{figure}

The last step was improving the performance for when the queries are received through the host, since it involves copying the queries to the device, processing them and then copying the results back to the host. The basic idea is to overlap the copying of queries with the processing of queries as much as possible. Figure \ref{fig:query_launch} illustrates this. In order to achieve this the array of queries is divided into chunks that get copied to the device sequentially, and once a chunk has been copied its processing can begin. Once the processing is done the results for that chunk can be copied back to the host. The ideal tool for this is CUDA Graphs\footnote{\url{https://docs.nvidia.com/cuda/cuda-runtime-api/group__CUDART__GRAPH.htm}}. CUDA Graphs allow the user to express the launch order and dependencies of different operations as a graph, which can then be launched as one object, allowing for ideal launch sequences. Creating a graph for the first time is expensive, but it can then be reused, since the parameters of the different operations of the graph can be changed. Figure \ref{fig:query_graph} shows a graph for the case where the queries are divided into four chunks. One important aspect of the graph is that the memory copies from the host to the device can never be more than one chunk ahead of the processing of the chunk. This reduces the space necessary on the device for the queries to only enough to hold two chunks. This means that if you have 100 queries divided into 10 chunks you only need two allocate enough space for 20 queries. The graph also enforces that the kernels are launched sequentially, in order to avoid the slowdown caused by having too many threads resident on the GPU.

\subsection{Rank Query}\label{rank_sec}
\begin{figure}[H]
    \centering
    \includesvg[width=.9\linewidth]{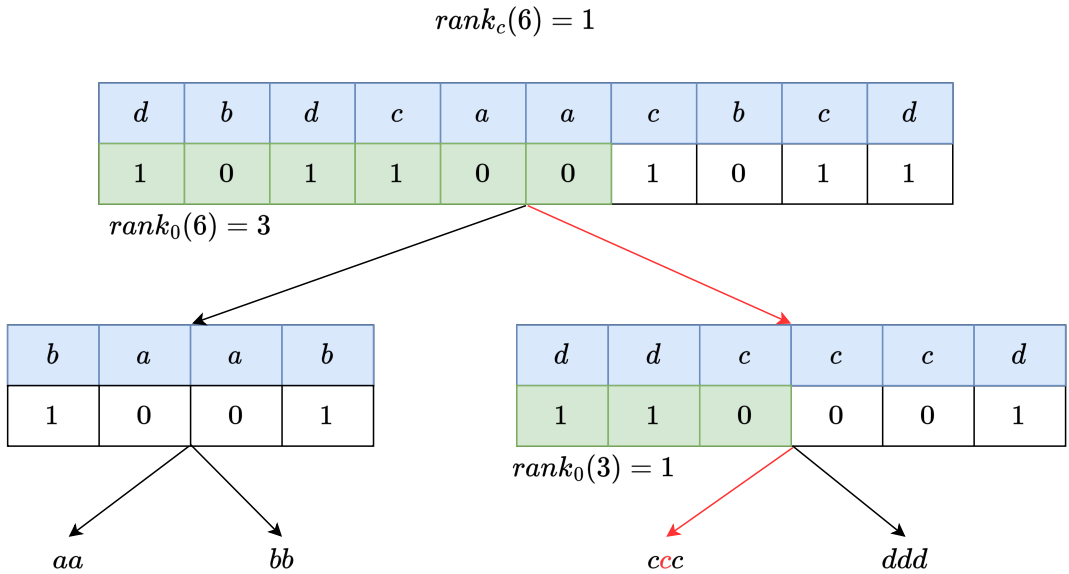}
    \caption{Example of querying the number of occurrences of symbol $c$ in the interval $[0, 6)$ using the wavelet tree of text $T = dbdcaacbcd$.}
    \label{fig:rank_query}
\end{figure}

The algorithm of the rank query is very similar to that of the access query. Algorithm \ref{alg:rank} illustrates it. The only differences with respect to the access query are that we now know the symbol, so we do not need to access the bit array at every level to determine whether to go left or right, and when we are at the leaf node, we return the position of the symbol rather than the symbol itself. Figure \ref{fig:rank_query} illustrates this. The query itself already gives us the symbol, so we know how to traverse the tree. The only information we need before traversing to the next level is how many bits we need to consider in our calculation. At the first level, since we know that we are traversing to the right child, we count the number of zeros before the position and subtract it from the position, which is equivalent to counting the number of ones before the position. With that we know how many bits we need to take into account at the next node, in this case three. We now do the same thing again, but since we are going left, we only need to count the number of zeros before the position, which in this case is one. Since we are at the end of the tree, the final result is one.

\begin{algorithm}[hbt!]
\caption{Wavelet tree rank query.}
\label{alg:rank}
\KwIn{Wavelet tree $wt$, index $i$ and symbol $c$ of desired rank.}
\KwOut{Number of occurences of $c$ in $[0, i)$.}
\Comment{Interval $[start\_symbol, end\_symbol)$ covered by current node.}
$start\_symbol \gets 0$\;
$end\_symbol \gets alphabet\_size$\;
$split\_symbol \gets 0$\;
$result \gets i$\;
\For{$l \gets 0$ \KwTo $wt.num\_levels$}{
\If{$end\_symbol - start\_symbol \leq 0$}{
    \textbf{break}\;
}
\Comment{Number of occurences of symbols smaller than $start\_symbol$. Corresponds to the start of the node in the level.} 
$counts \gets wt.cum\_histogram[start\_symbol]$\;

\Comment{Number of zeros until start of node.}
$start \gets wt.rank_0(l, counts)$ \;
\Comment{Number of zeros until desired index, $result$ is relative to the start of the node.}
$pos \gets wt.rank_0(l, counts + result)$ \;
$split\_symbol \gets start\_symbol +  prevPowTwo(end\_symbol - start\_symbol)$\;
\If{$c < split\_symbol$}{
    $result \gets pos-start$\;
    $end\_symbol \gets split\_symbol$\;
}
\Else{
    $result \gets result - (pos-start)$\;
    $start\_symbol \gets split\_symbol$\;
}
}
\Return $result$\;
\end{algorithm}

As with the access query, in the implementation, the binary rank in line 10 is pre-computed.

\subsubsection{Rank query optimization}
Being so similar to the access query, the rank query benefits from all the optimizations covered before. Additionally, the same CUDA graph can be used to launch rank queries from the host, since the exact kernel that will be launched can be changed after instantiating the graph.

\subsection{Select query}\label{select_sec}
\begin{figure}[H]
    \centering
    \includesvg[width=.9\linewidth]{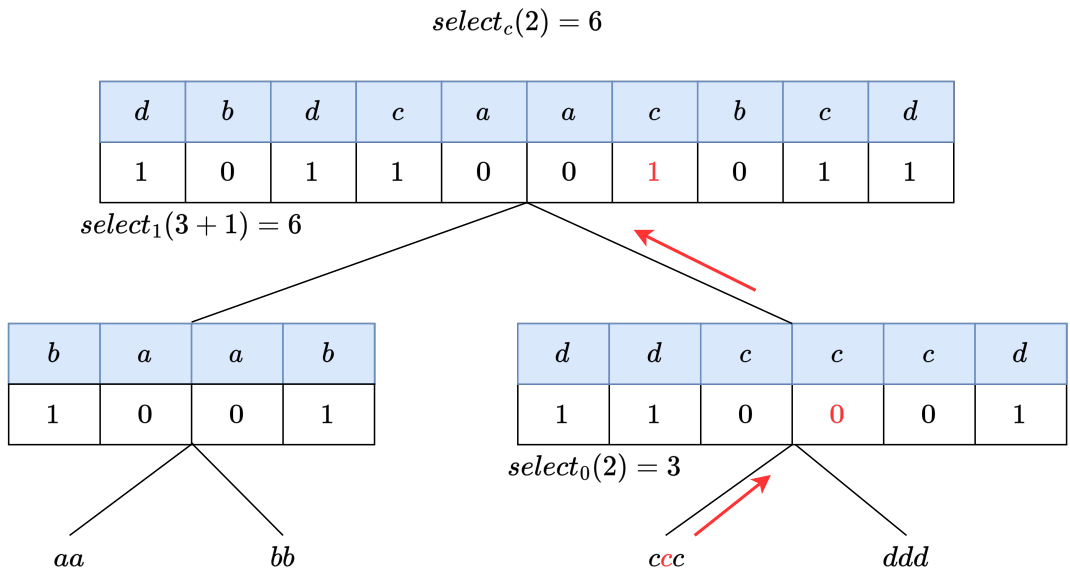}
    \caption{Example of querying the position of the second $c$ using the wavelet tree of text $T = dbdcaacbcd$.}
    \label{fig:select_query}
\end{figure}

The select query is quite different from the other two. Since we want to find the position of a specific occurrence of a symbol, we need to start at a leaf node and then traverse upwards until we reach the root node. Algorithm \ref{alg:select} and  Figure \ref{fig:select_query} illustrate this process. As with the rank query, we already know how to traverse the tree, since the symbol is given in the query itself. Since we are looking for the second occurrence of the symbol $c$, we will perform a $select_0(2)$ to get its position in the right node of the second level, which is our starting point. Since in the current node the desired occurrence is at position three, we know that in the parent node it will be the fourth one-bit. To find the position of the fourth one-bit, we do a $select_1(4)$. If the tree had more levels we would continue as we have until now, but since we have reached the root node, we are done.

\begin{algorithm}[hbt!]
\caption{Wavelet tree select query. Inner workings of $getNodeStart$ function on line 5 are shown in Algorithm \ref{alg:getNodePos}.}
\label{alg:select}
\KwIn{Wavelet tree $wt$, index $i$, starting from $1$, and symbol $c$.}
\KwOut{Position of the $i-th$ occurrence of $c$.}
\Comment{Interval $[start\_symbol, end\_symbol)$ covered by current node.}
$start\_symbol \gets c$\;
$encoded\_symbol \gets encode(c)$\;
$result \gets i$\;
\For{$l \gets encoded\_symbol.length - 1$ \KwTo $0$}{
\Comment{If current node is right child of parent.}
\If{$getBit(encoded\_symbol.code,wt.num\_levels - 1 -l)  == 1$}{
    $start\_symbol \gets getNodeStart(start\_symbol, l, num\_levels)$\;
    \Comment{Number of occurences of symbols smaller than $start\_symbol$. Corresponds to the start of the node in the level.} 
    $counts \gets wt.cum\_histogram[start\_symbol]$\;

    \Comment{Number of ones until start of node.}
    $start \gets wt.rank_1(l, counts)$\;
    $result \gets wt.select_1(l, start + result) +1$\ \Comment*{1-indexed.}
}
\Else{
    $counts \gets wt.cum\_histogram[start\_symbol]$\;

    \Comment{Number of zeros until start of node.}
    $start \gets wt.rank_0(l, counts)$\;
    $result \gets wt.select_0(l, start + result) +1$\ \Comment*{1-indexed.}
}
\Comment{Make result relative to start of node.}
$result \gets result - counts$\;
}
\Return $result-1$\ \Comment*{0-indexed.}
\end{algorithm}

\begin{algorithm}[hbt!]
\caption{Algorithm to find at which symbol a wavelet tree node starts given a symbol it covers and its level.}
\label{alg:getNodePos}
\KwIn{Symbol $c$, level $l$ and wavelet tree $wt$.}
\KwOut{Symbol at which the node starts.}
$node\_start \gets 0$\;
\If{$isPowTwo(wt.alphabet\_size)$}{
    $node\_len\_at\_level \gets 1 \ll (wt.num\_levels - l)$\;
    $node\_start \gets c\ \&~(node\_len\_at\_level - 1)$\;
}
\Else{
    $code \gets wt.encode(symbol)$\;
    $is\_rightmost\_node \gets popcount(code.code \gg (wt.num\_levels - l)) == l$\;
    \If{$is\_rightmost\_node$}{
        \For{$i \gets 0$ \KwTo $l$}{
            $node\_start \gets node\_start + getPrevPowTwo(wt.alphabet\_size-node\_start)$\;
        }
    }
    \Else{
        $node\_start \gets c - 2^{code.len - l -1}$\;
    }
}
\Return $node\_start$

\end{algorithm}

As with the other queries, the binary ranks on lines 8 and 13 are precomputed. In this instance the optimization improved runtime on an NVIDIA RTX 2080 Ti by 7\% to 23\%, the lower bound corresponding to an alphabet size $\sigma = 4$, and the upper one to $\sigma = 20'000$. I also added the pre-computed binary ranks to shared memory whenever they fit.

\subsubsection{Select query optimization}
Since most of the computation is spent on the binary select queries at every level, the optimization focused on that part of the implementation. The binary select query can be broken down into three parts. First, we must find the L1-block where the desired bit is located. We then have to find the correct L2-block. Finally, we must go through all the words in the L2-block until we find the desired bit. 

As with the other queries, I implemented the option of having different numbers of threads per query (TPQ). In the initial implementation, that differed from Algorithm \ref{alg:bin_select} in that the searches were linear and not binary, the one TPQ configuration was the fastest when profiled on an NVIDIA RTX 2080 Ti. It was around four times faster than the 32 TPQ configuration, which was around three times faster than all others. Looking at the performance metrics, it was clear that the performance degradation of the other TPQ configurations was due to a large amount of warp stalling when communicating between threads in the group. When profiled on an NVIDIA A100, the results were very different. The one TPQ configuration was still the fastest, but only around two times faster than the runner-up, which was the 2 TPQ configuration, and the runtime increased monotonically with increasing numbers of threads per query. When looking at the performance metrics, the warp stalling when communicating between threads was no longer there.

Switching to a binary search when searching for the correct L2-block is ideal, since all L1-blocks except the last will be complete, which in this case means having 128 L2-blocks. For the configurations with multiple TPQ, the binary search was implemented such that the search space was sliced into chunks and each thread in the group performed its own binary search on its own chunk of the search space. The threads then had to communicate with one another to get the correct L2-block from the thread that found it. All configurations improved with this change, but the one TPQ configuration was still the fastest.

For the first part, the main optimization decision is also which searching algorithm to use. Since the size of an L1-block is 65'536 bits, and the 16'384-th one and zero bit is sampled, in most cases the search space given by the samples before and after the desired bit will be only one or two L1-blocks. This means that linear search should be more efficient in these cases. On the other hand, linear search would be inefficient for the edge cases where the search space is very large. Comparing linear and binary search on uniformly randomly distributed texts, binary search is 3\% to 8\% slower than linear. This is a small drawback compared to the robustness of binary search on any search space size. For this reason I decided to stick with binary search.

Finally, the last part of the optimization focused on the loop (lines 22-30 in Algorithm \ref{alg:bin_select}). In the implementation, when there is more than one TPQ, each thread in the group accesses an adjacent word of the bit array, resulting in a coalesced global memory access. The threads then calculate the number of ones (or zeros) in the word and perform an inclusive prefix sum, so that each thread has the number of ones (or zeros) up to and including its own word. With that information each thread can figure out whether the desired bit is in its own word, and then find out exactly where it is located. At the end of every iteration the threads need to communicate with one another to know whether one of them has found the desired bit and the work is done. The inclusive sum and communication are efficient since they benefit from warp intrinsics, but it is still extra computation that the one TPQ configuration does not have to do. As with the loop in the binary rank operation, I made the access to the bit array two words (or one 64-bit word) at a time. This improved runtime for the one TPQ configuration by around 11\%. All TPQ configurations except the 16 and 32 TPQ benefited from this change. This is logical since when accessing two words at a time the maximum number of bit array accesses given an L2-block size of 512 bits will be eight, so most of the threads in the group will not be contributing.

An important part of the loop is finding the exact position of the desired bit inside a word once the desired word has been found. Initially, I implemented this operation such that if we want to find the $n^{th}$ set bit, we repeatedly remove the least significant set bit from the word $n-1$ times, and then extract the position of the least significant bit of the word using the math intrinsic $\_\_ffs$. This is efficient since at every iteration we are only performing an integer addition and a bitwise and operation, but we still have to iterate $n-1$ times. An alternative implementation is to perform a binary search, where you divide the word into two and perform a $popcount$ on both halves in order to find out which of them contains the desired one (or zero), and proceed recursively until only one bit is left. This requires five iterations for a 32-bit word and six for a 64-bit word. When profiled on an RTX 2080 Ti, this change improved runtime by 2\%, reducing the amount of executed instructions by 25\% on average, but increasing warp stalling by 30\%. On an A100 however, the improvement was between 6\% and 14\%, due to a similar reduction in executed instructions with a smaller increase in warp stalling.

As with the other queries, the same CUDA graph can be used to launch select queries from the host, since the exact kernel that will be launched can be changed after instantiating the graph.

\newpage\null\thispagestyle{empty}\newpage
\section{Experimental Results}
\subsection{Experimental Setup}
I conducted all the experiments of the CPU libraries using a server node containing a 32-core Ryzen Threadripper 3970X with hyperthreading enabled, and 256 GB of DDR4 RAM with a clock rate of 3200 MHz, from now on referred to as "CPU". The GPU experiments of this library were performed on two different systems:

\begin{itemize}
    \item An Intel Xeon Gold 6226R, with two sockets, 16 cores per socket and 64 hyperthreads, an NVIDIA RTX 3090 GPU, and 64GB of DDR4 RAM with a clock rate of 2933 MHz. From now on referred to as "3090".
    \item An Intel Xeon Gold 6226R, with two sockets, 16 cores per socket and 64 hyperthreads, an NVIDIA A100 40GB PCIe GPU, and 64GB of DDR4 RAM with a clock rate of 2933 MHz. From now on referred to as "A100".
    %\item A Ryzen Threadripper 2950X, with 16 cores and 32 hyperthreads, an NVIDIA RTX 4090 GPU, and 48GB of DDR4 RAM with a clock rate of 2133 MHz. From now on referred to as "RTX 4090".
\end{itemize}

The benchmarks were performed using either the Google Benchmark\footnote{\url{https://github.com/google/benchmark}} C++ library or a custom benchmarking script, where I made sure that all the runs had at least 10 iterations, and selected the median time. For CUDA benchmarks, I included a warm-up run at the beginning of a program in order to initialize the CUDA context.

For the benchmarks of the rank and select structure I used the bit arrays described by Kurpicz \cite{Kurpicz_RS}. In their experiments, they create random bit arrays with three different fill rates (10\%, 50\% and 90\%  of all bits are ones), and differentiate between two distributions. One is a uniform distribution, where the one bits are evenly distributed across the whole bit array. The other is an adversarial distribution, which they define as: "\textit{Assume that k\% of the bits in the bit vector should be ones. Then, we set 99\% of the ones in the last k\% of the bit vector and the remaining one percent in the first 100 - k\% of the bit vector.}"

For the benchmarks of the wavelet tree, I use the \textit{CommonCrawl}, \textit{DNA} and \textit{Protein} texts as described by Dinklage et al. \cite{practical_WT}. The exact script that prepares these texts is in the GitHub repository of the project. Table \ref{tab:text_characteristics}, adapted from \cite{practical_WT}, shows the characteristics of each text. The empirical entropy~\cite{entropy} measures the compressibility of a sequence (in this case a text) based on the probability distribution of its symbols or sequences of symbols. The order of the entropy describes how many previous symbols are taken into account when calculating the probability of a symbol appearing.

\begin{table}[h]
    \centering
    \caption{Characteristics of the Texts Used for the Wavelet Tree Benchmarks: Name of the Text, Alphabet Size $\sigma$ of the Whole Text, Alphabet Size $\sigma_x$ for an $x$-character Prefix of the Text, Total Text Size $n$, and Empirical Entropy $H_k$ for $k \in [0,3]$}
    \label{tab:text_characteristics}
    \begin{tabular}{l c c c c r c c c c}
        \hline
        Name & $\sigma$ & $\sigma_{2^{30}}$ & $\sigma_{2^{31}}$ & $\sigma_{2^{32}}$ & $n$ & $H_0$ & $H_1$ & $H_2$ & $H_3$ \\
        \hline
        CommonCrawl & 243 & 243 & 243 & 243 & 196,885,192,752 & 6.19 & 4.49 & 2.52 & 2.08 \\
        DNA & 4 & 4 & 4 & 4 & 218,281,833,486 & 1.99 & 1.97 & 1.96 & 1.95 \\
        Prot & 25 & 25 & 25 & 25 & 50,143,206,617 & 4.21 & 4.20 & 4.19 & 4.17 \\
        \hline
    \end{tabular}
\end{table}

I also benchmarked the wavelet tree implementation using uniformly randomly distributed texts for a variety of alphabet sizes to get a more thorough comparison.

\subsection{Rank and Select structure}
Since the implementation is very similar to \textit{wide-popcount} from \cite{Kurpicz_RS}, I will be using that as the benchmark CPU implementation. It is also the most performant implementation on rank queries for its amount of overhead, but the author deems it unsuited for select queries. For the configuration of the structures themselves I used an L2-block size of 512 bits and a select sample rate of 16'384, leading to an overhead of 3.6\%. In Table 1 of \cite{Kurpicz_RS}, the overhead of \textit{wide-popcount} is said to be around 10\%, which does not make sense given its description. In any case, the overhead of my implementation is similar to the ones with the lowest overhead.

Since the GPU is a parallel throughput machine, when comparing the performance of queries I believe it is best to compare the throughput. It was calculated by measuring the time taken to process 100'000'000 randomly distributed queries. In order to have as fair a comparison as possible, when performing $rank_0$ or $select_0$ queries I specialized the CPU implementation on zero queries, and one queries otherwise, since it was available as an option. This specialization dictates whether zero or one bits are counted for the rank structures. I also computed the queries in parallel using the OpenMP $parallel\ for$ directive. For my implementation, I launched as many threads as necessary in order to fully occupy the GPU, and processed queries in a grid-strided loop. I used the one TPQ configuration for both rank and select as it is the fastest.

\subsubsection{Construction}
\begin{figure}[H]
    \centering
    \includegraphics[width=.9\linewidth]{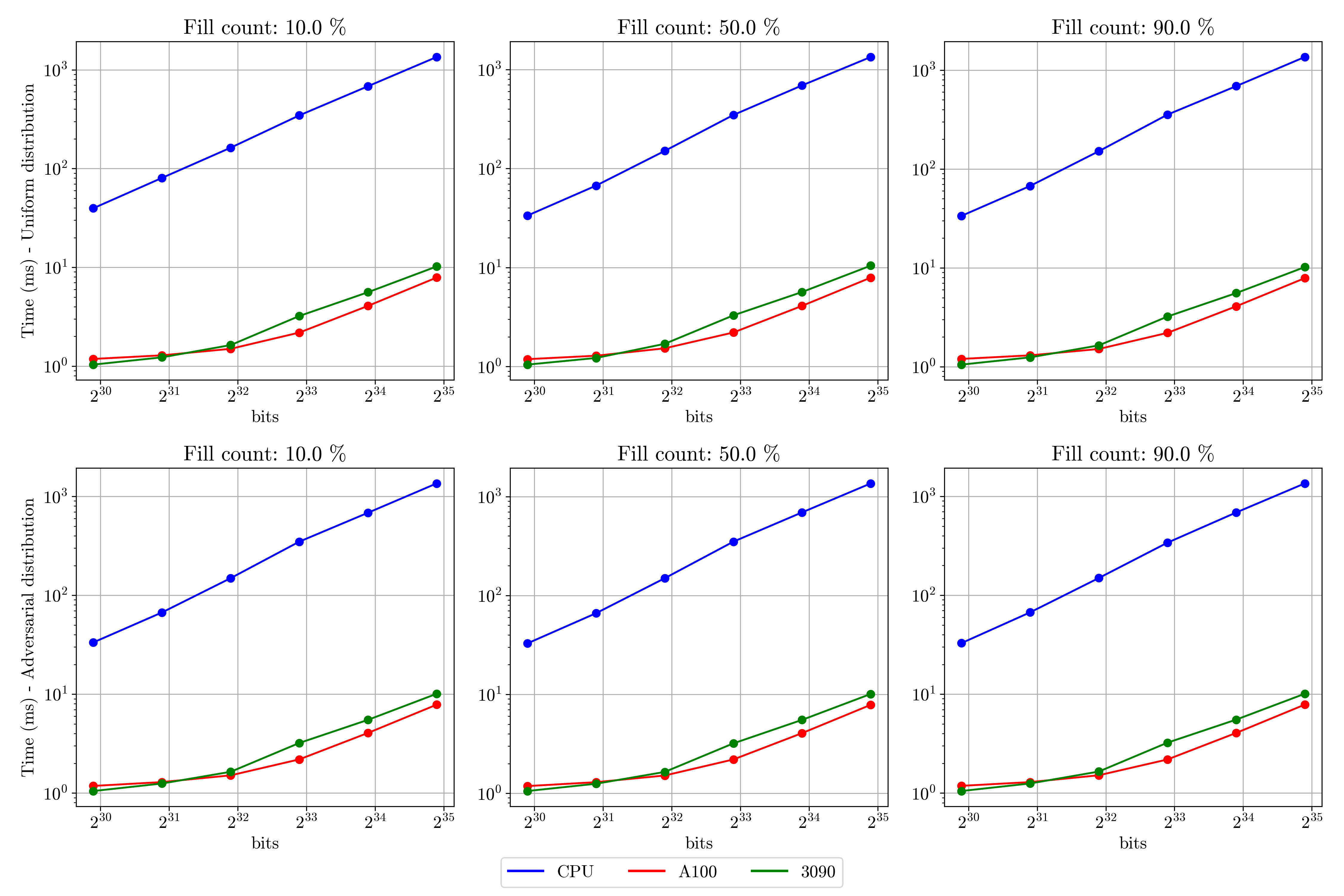}
    \caption{Experimental results of rank and select support structure construction time.}
    \label{fig:RS_construction}
\end{figure}

When comparing the CPU and my own construction algorithms, it is important to note that the CPU construction algorithm is sequential, while my own benefits from the high amount of parallelism that a GPU offers. Figure \ref{fig:RS_construction} shows construction times for different bit sizes depending on the fill count and whether the distribution of ones is uniform or adversarial. An important characteristic that the figure illustrates is that the construction time is completely independent from the content of a bit array. For the smaller bit array sizes, the GPU implementation's time barely increases, which is most probably due to the GPU not being fully utilized. Overall, the GPU implmentation is between 40 and 130 times faster than the CPU one.

\subsubsection{Binary Rank}
\begin{figure}[H]
    \centering
    \includegraphics[width=.9\linewidth]{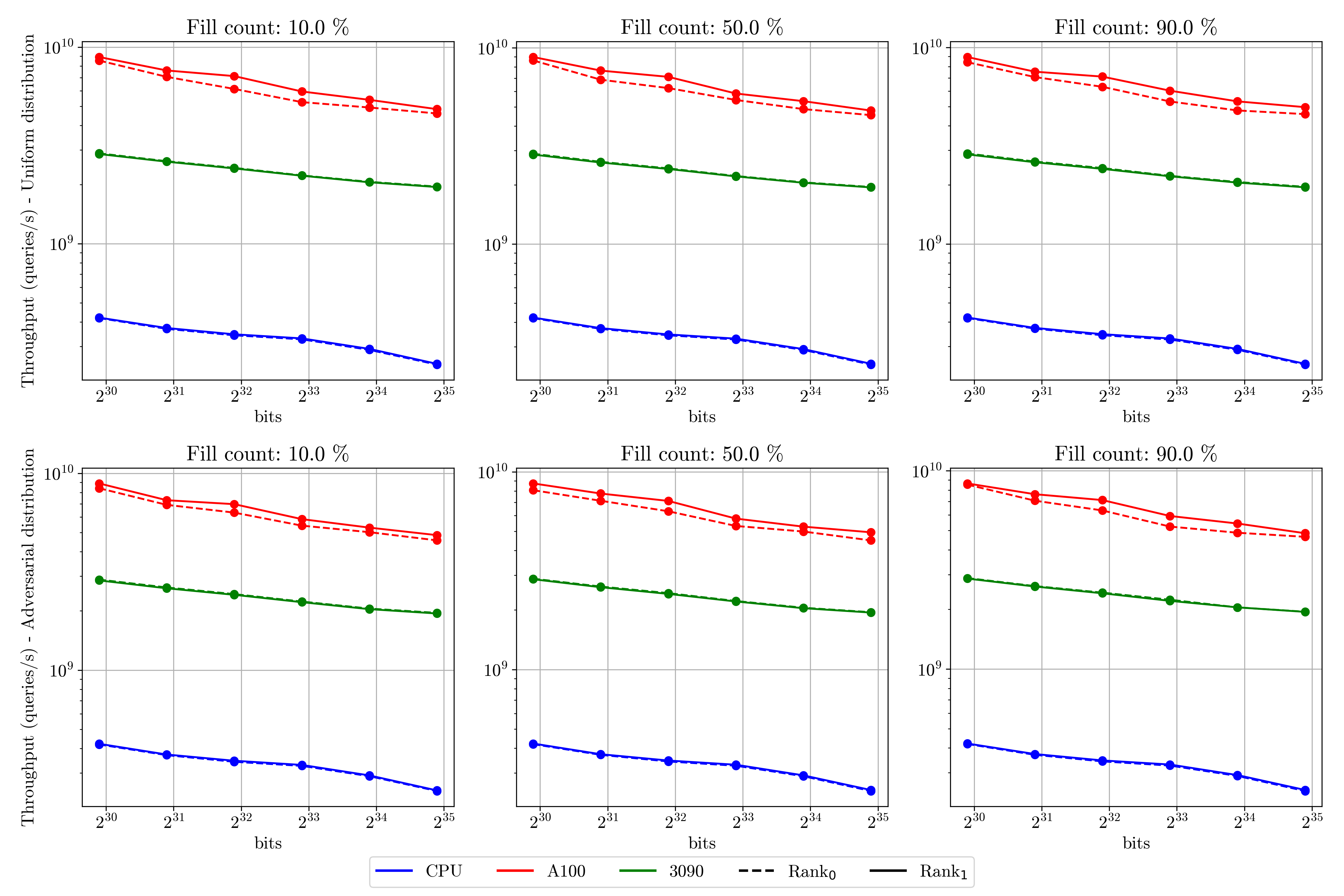}
    \caption{Experimental results of binary $rank_0$ and $rank_1$ queries.}
    \label{fig:RS_rank}
\end{figure}

 Figure \ref{fig:RS_rank} shows the results for $rank_0$ and $rank_1$ queries. From the figure we can see that the throughput of the 3090 is around 10 times higher than that of the CPU implementation, while that of the A100 is even higher, around 22 times. In all cases, the throughput decreases with increasing bit array size, which is to be expected due to decreasing benefit from caching. Another positive characteristic of all implementations is that their throughput is independent from the content of the bit array. Interestingly, $rank_1$ queries have slightly higher throughput than $rank_0$ queries on the A100, even though the only difference in the implementation is that the total number of ones until the position has to be subtracted from the position itself to get the total number of zeros.

\subsubsection{Binary Select}
Since the authors in \cite{Kurpicz_RS} deemed the \textit{wide-popcount} (CPU-wide) implementation unsuitable for select queries, I also include the \textit{flat-popcount} (CPU-flat) one, which is the most performant of the low-overhead rank and select structures on select queries. Even though the most performant version of it is the one that uses SIMD intrinsics to find the desired L2-block, I used the binary search option, as the other did not compile correctly.

\begin{figure}[H]
    \centering
    \includegraphics[width=.9\linewidth]{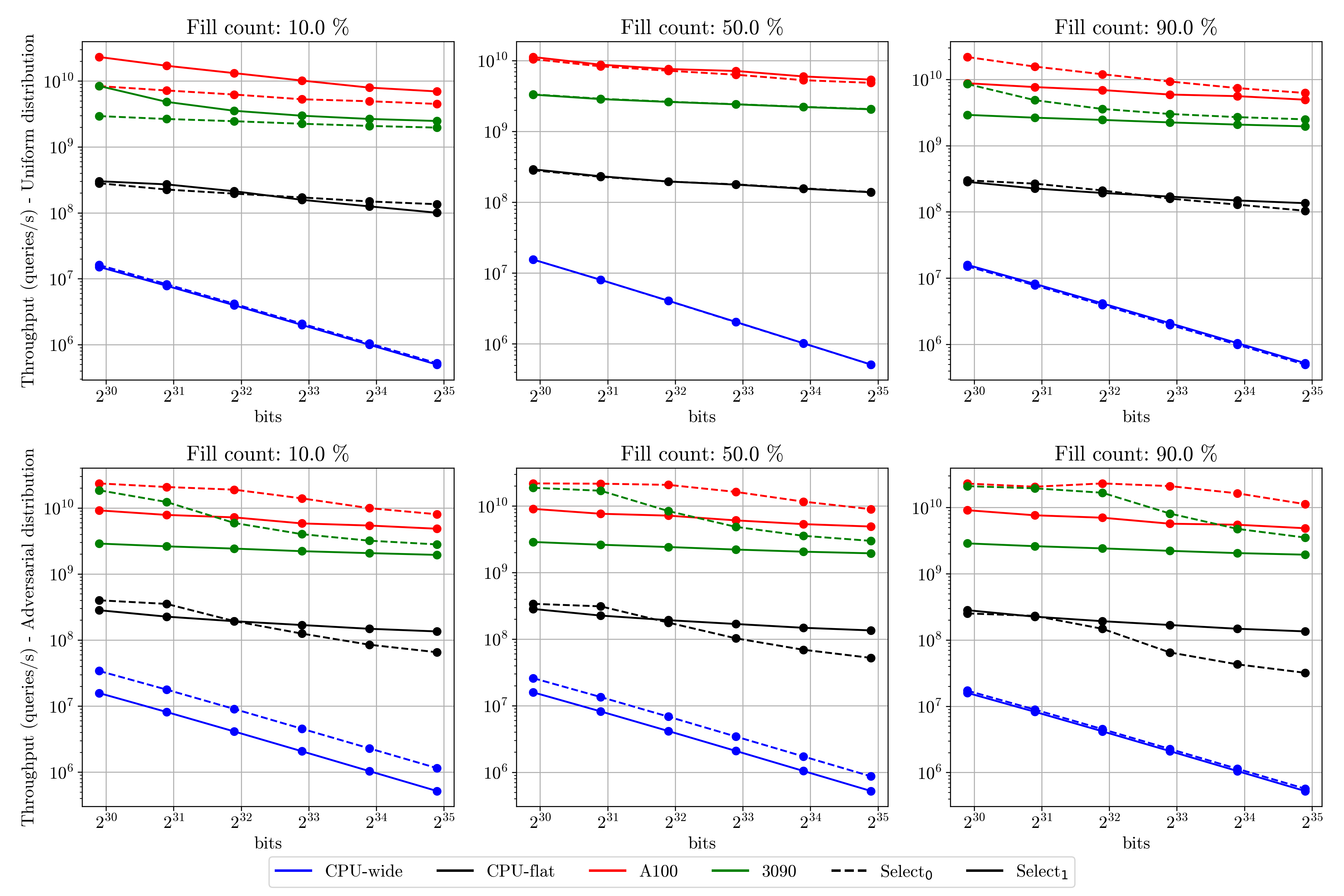}
    \caption{Experimental results of binary $select_0$ and $select_1$ queries.}
    \label{fig:RS_select}
\end{figure}

Figure \ref{fig:RS_select} shows the results for $select_0$ and $select_1$ queries. From the figure we can see that the throughput of the 3090 is at least 100 times higher than that of the CPU-wide implementation, becoming up to around 3'700 times higher in some cases. The A100's throughput is between two and three times higher than 3090. In all cases the throughput decreases with increasing bit array size, which is to be expected due to decreasing benefit from caching, but the throughput of the CPU-wide implementation decreases substantially faster. Compared to CPU-flat, the 3090 achieves at least 10 times higher throughput.

For the uniform distribution, having a low percentage of the value that we are looking for (zeros for $select_0$ and vice-versa), results in higher throughput. The reason for this could be that having a bigger search space results in that more of the L1-index is loaded into the cache, which benefits all threads resident in an SM. Also, for the adversarial distribution, the $select_0$ query has always higher throughput than the $select_1$ query. This is to be expected, as the one bits are adversarially distributed, and not the zero bits, so the distribution of zero bits is slightly more favorable. Interestingly, CPU-flat exhibits the opposite behavior.

\subsection{Wavelet Tree}
For the construction of the wavelet tree, I compare my own implementation against the best algorithms from \cite{practical_WT}. They also use approximately the same amount of extra space as mine. These are the $par.dd.pc$ and $par.dd.pc.ss$ algorithms. Since in their benchmarking script the authors do not include the construction of the rank and select support structures, I don't either. They also convert the text to a minimal alphabet before constructing the tree, which I do, too.

For the queries, I compare against the implementation from the Succint Data Structures Library (SDSL), specifically against the balanced tree with $rank\_support\_v5$ rank and $select\_support\_mcl$ select support. The rank support has an overhead of 6.25\%. Since the overhead of the select support structure is dependent on the number of ones in the bit array, it is hard to give a value for it, but Kurpicz (\cite{Kurpicz_RS} Table 1) found the average overhead to be around 18.5\%. For my implementation, I use an L2 block size of 512 bits and sample every 4'096-th one and zero bit, resulting in a maximal overhead of $4.8\%$. When processing the queries using the SDSL wavelet tree, I parallelize the loop with the OpenMP $parallel\ for$ directive, in order to have as fair a comparison as possible. The timing of my implementation also includes the time taken to copy the queries from the host to the device, and the results from the device back to the host. As mentioned in Section \ref{access_sec}, for processing the queries I use CUDA Graphs. The usage of a copy node requires that the host memory be pinned, so one factor that makes a big difference in the time it takes to process queries is whether they have been allocated using a pinned memory allocator or not. When comparing the SDSL library with my own, I will always use the case where the queries are not pinned and later show the performance improvement of allocating the queries using pinned memory. It is important to note that in the comparison the additional time necessary to allocate pinned memory instead of pageable memory is not included. To see how the implementations compare for different numbers of queries, instead of showing the throughput, I will show the time taken to process different numbers of queries. This should also give the user an idea of how many queries need to be processed in parallel so that it is worth using the GPU.

\subsubsection{Construction}
First we will perform the comparison on uniformly randomly distributed texts for a variety of alphabet sizes in order to get a more thorough comparison. After that we will use real-world texts.

\begin{figure}[H]
    \centering
    \includesvg[width=.9\linewidth]{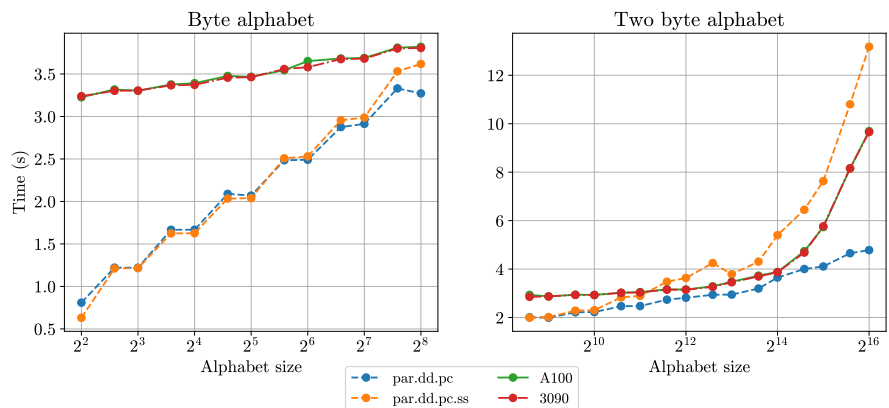}
    \caption{Construction time of my own wavelet tree and the best parallel CPU implementations on uniformly randomly distributed texts of 8 GB size for a variety of alphabet sizes.}
    \label{fig:wt_constr_random}
\end{figure}

Figure \ref{fig:wt_constr_random} compares the construction times for uniformly randomly distributed texts of 8 GB size for a variety of alphabet sizes. Overall, my implementation does not manage to surpass the performance of the state of the art. One interesting thing that can be seen in the figure is that for smaller alphabet sizes the construction time increases marginally as the alphabet size increases. This is most probably because the construction time is dominated by the text allocation and memory copy, which is independent of the alphabet size. Also, my algorithm and $par.dd.pc.ss$ exhibit a substantial increase in construction time for $\sigma \geq 2{14}$, while $par.dd.pc$ continues to scale linearly. In my implementation, the cause for this increase is the extraction of the alphabet from the text. When investigating why $par.dd.pc$ did not exhibit such an increase, I found that for their algorithms, the authors in~\cite{practical_WT} pass the number of levels of the wavelet tree as an argument. Since they also first convert the text to a minimal alphabet, which I do as well for fairness, this allows them to construct the tree without having to extract the alphabet from the text. I therefore believe it is fairer to pass the alphabet of the text as an argument to my wavelet tree, which is a feature that is supported by the API. The results of this comparison I will discuss shortly. Lastly, the figure also shows how the extra computation necessary to build the reduced tree in case the alphabet size is not a power of two affects the construction time only minimally.

\begin{figure}[H]
    \centering
    \includesvg[width=.9\linewidth]{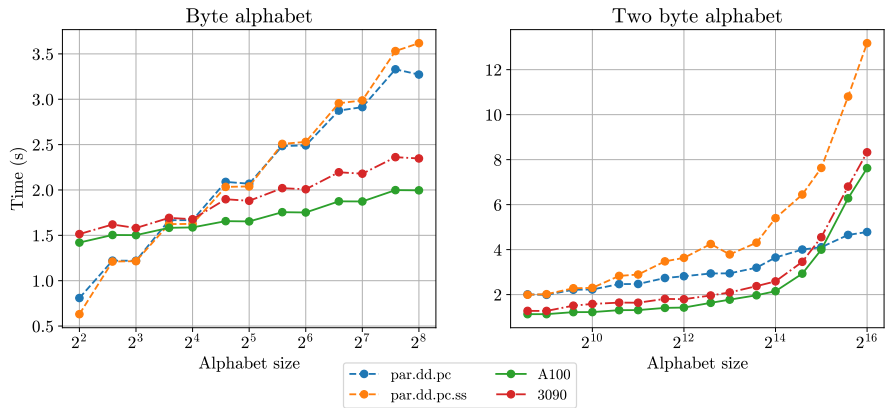}
    \caption{Construction time of my own wavelet tree, without accounting for allocation and copy time, and the best parallel CPU implementations on uniformly randomly distributed texts of 8 GB size for a variety of alphabet sizes.}
    \label{fig:wt_constr_random_no_copy}
\end{figure}

Figure \ref{fig:wt_constr_random_no_copy} compares construction times without taking into account the time necessary to allocate memory for the text on the GPU and copy it from the CPU to the GPU. In this case, the A100 becomes faster than the CPU algorithms from $\sigma \geq 12$, and the 3090 from $\sigma > 16$. However, due to the quadratic behavior of my algorithm from $\sigma \geq 2^{14}$ onwards, $par.dd.pc$ becomes faster for $\sigma \geq 2^{15}$. Interestingly, for the byte alphabet my algorithm still scales better with the alphabet size. Now that the copy time and allocation time is removed, there is also a difference in construction time between the 3090 and A100, the A100 always being faster.

\begin{figure}[H]
    \centering
    \includesvg[width=.9\linewidth]{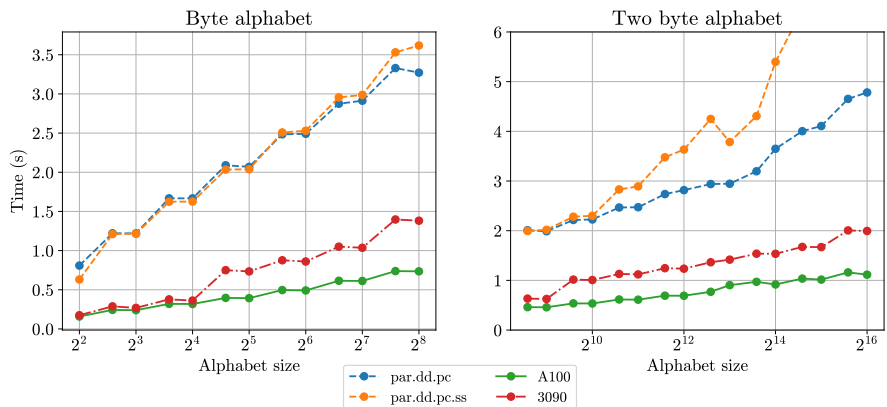}
    \caption{Construction time of my own wavelet tree, without accounting for allocation and copy time, and the best parallel CPU implementations on uniformly randomly distributed texts of 8 GB size for a variety of alphabet sizes. The alphabet of the text is passed as an argument in my implementation.}
    \label{fig:wt_constr_random_no_copy_alphabet}
\end{figure}

Figure \ref{fig:wt_constr_random_no_copy_alphabet} compares construction times without allocation and copy time, as well as passing the alphabet as an argument to my wavelet tree. Here we can see that my implementation is always faster and scales better with the alphabet size than the others. The 3090 is between 2x and 4.5x faster than the best option from the CPU construction algorithms. The A100 is between 3.2x and 5.2x faster. When comparing the GPUs, the A100 is always faster, with the difference increasing with the alphabet size for both the byte and the two-byte alphabet.

\begin{figure}[H]
    \centering
    \includesvg[width=.9\linewidth]{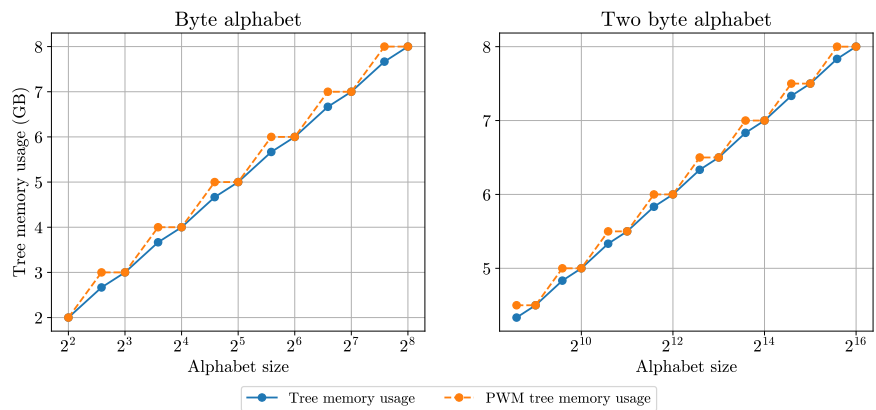}
    \caption{Memory usage of my own wavelet tree and the tree created by the best parallel CPU implementations on uniformly randomly distributed texts of 8 GB size for a variety of alphabet sizes.}
    \label{fig:wt_constr_random_mem}
\end{figure}

\begin{figure}[H]
    \centering
    \includesvg[width=.9\linewidth]{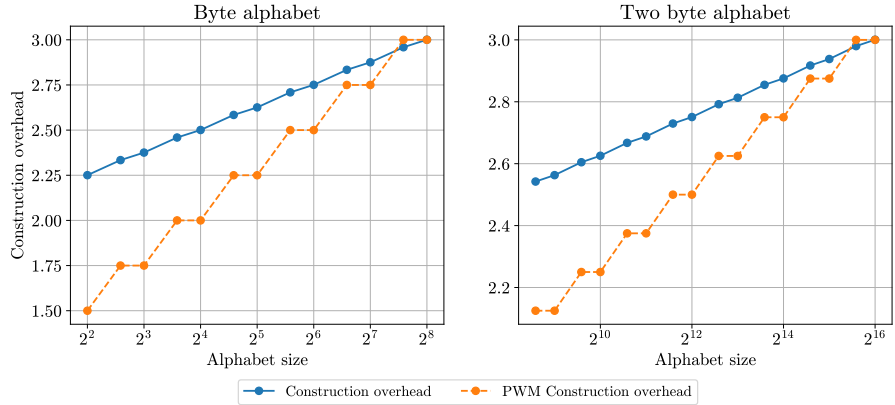}
    \caption{Wavelet tree maximum construction overhead comparison between my own construction algorithm and the best CPU parallel construction algorithms on uniformly randomly distributed texts of 8 GB size for a variety of alphabet sizes. Overhead is calculated relative to the text size, an overhead of 1 would mean that no extra space is used.}
    \label{fig:wt_constr_random_overhead}
\end{figure}

Figure \ref{fig:wt_constr_random_mem} shows the size of the wavelet tree created, and Figure \ref{fig:wt_constr_random_overhead} shows the amount of overhead compared to the original text necessary for constructing the tree, for the same uniformly randomly distributed texts. When comparing the space necessary for the tree, we can see that for alphabet sizes that are not a power of two my own tree occupies less space. The difference in space usage will probably be less in real-world texts which are not uniformly distributed. The difference in construction overhead comes from how the CPU implementation works. Each thread constructs a partial wavelet tree on a portion of the text, which is then merged into a global tree. The sequential construction algorithm used for computing the partial wavelet trees only requires $\mathcal{O}(\sigma)$ additional space. In my algorithm, additional space is necessary for the sorted text, which must be of the same data type as the original text, due to how the CUB API is structured. Having a custom radix sort implementation would allow for the sorted text to have only as many bits per character as necessary, which would bring the construction overhead down to a similar level than that of the CPU implementation.

\begin{figure}[H]
    \centering
    \includesvg[width=.9\linewidth]{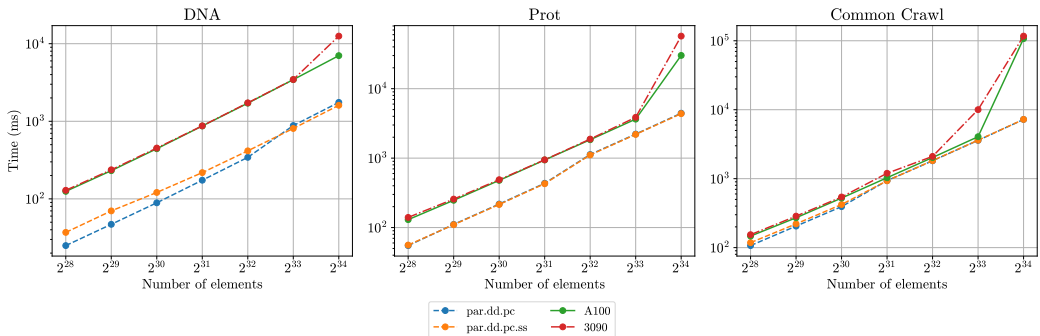}
    \caption{Construction time of my own wavelet tree and the best parallel CPU implementations on real-world texts.}
    \label{fig:wt_constr_data}
\end{figure}

Now we will compare the implementations on real-world texts. Figure \ref{fig:wt_constr_data} shows the difference in construction time. At first glance we can already see that both algorithms scale similarly with increasing data size, except when the size becomes too large and unified memory has to be used, which causes a stark increase in construction time. When we compare the construction time across alphabet sizes, we can see that the construction time of my own implementation barely increases, while that of the CPU implementations edges closer. This is consistent with what we saw on Figure \ref{fig:wt_constr_random}.

\begin{figure}[H]
    \centering
    \includesvg[width=.9\linewidth]{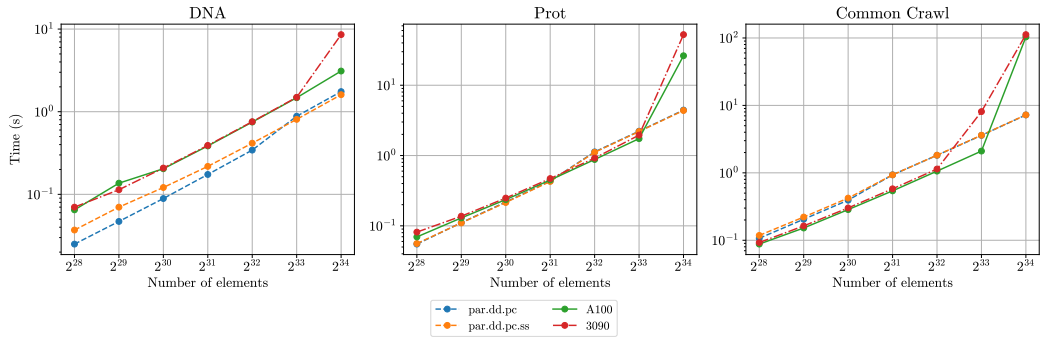}
    \caption{Construction time of my own wavelet tree, without accounting for allocation and copy time, and the best parallel CPU implementations on real-world texts.}
    \label{fig:wt_constr_data_no_copy}
\end{figure}

Figure \ref{fig:wt_constr_data_no_copy} compares construction times without taking into account the time necessary to allocate memory for the text on the GPU and copy it from the CPU to the GPU. On DNA, both the A100 and 3090 are between 3x and 1.7x slower, the difference decreasing as the text size increases. On Prot, the A100 is at most 1.25x slower, and the 3090 1.5x slower. For bigger text sizes they become up to 1.25x faster. On Common Crawl, the A100 is between 1.2x and 1.7x faster, while the 3090 is between 1.2x and 1.6x faster.

Figure \ref{fig:wt_constr_random_no_copy_alphabet} compares construction times without allocation and copy time, as well as passing the alphabet as an argument to my wavelet tree. The following speedup comparisons take into account only data sizes that do not need unified memory to be used, since that severely slows down the construction. On DNA, both the A100 and 3090 are between 2x and 4.5x faster, the difference increasing with the text size. On Prot, the A100 is between 3x and 5x faster, while the 3090 is between 2.4x and 4.2x faster. On Common Crawl, the A100 is between 3.5x and 4.7x faster, while the 3090 is between 2.8x and 3.8x faster. The A100 is always faster than the 3090.

\begin{figure}[H]
    \centering
    \includesvg[width=.9\linewidth]{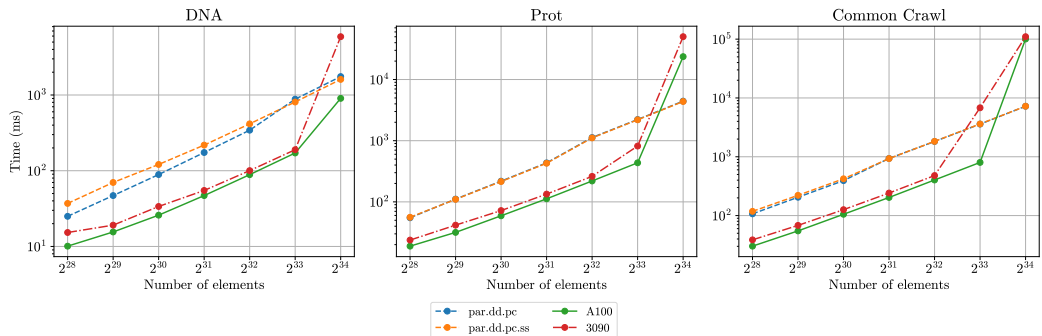}
    \caption{Construction time of my own wavelet tree, without accounting for allocation and copy time, and the best parallel CPU implementations on real-world texts. The alphabet of the text is passed as an argument in my implementation.}
    \label{fig:wt_constr_data_no_copy_alphabet}
\end{figure}

\begin{figure}[H]
    \centering
    \includesvg[width=.9\linewidth]{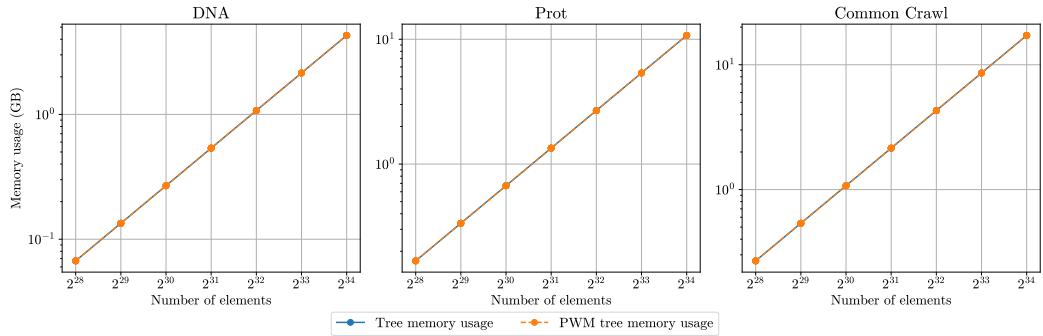}
    \caption{Memory usage of my own wavelet tree and the best parallel CPU implementations on real-world texts.}
    \label{fig:wt_constr_data_mem}
\end{figure}

Looking at the space the constructed wavelet tree uses, shown in Figure \ref{fig:wt_constr_data_mem}, we see almost no difference. For DNA, since $\sigma = 4$, which is a power of two, there can be no advantage gained from encoding the alphabet. For Prot, with $\sigma = 25$, only one symbol is encoded, which happens to be one that seldom appears in the text, leading to imperceptible savings in space usage. For Common Crawl, with $\sigma = 243$, three symbols are encoded, which again are ones that seldom appear in the text.

Overall, there is room for improvement. The areas with the most potential are mitigating the overhead of allocating and copying the data to the GPU and implementing a custom radix sorting algorithm that allows for the sorted text to be allocated using only the number of bits per symbol that are necessary.

\subsubsection{Access Query}
\begin{figure}[H]
    \centering
    \includesvg[width=.9\linewidth]{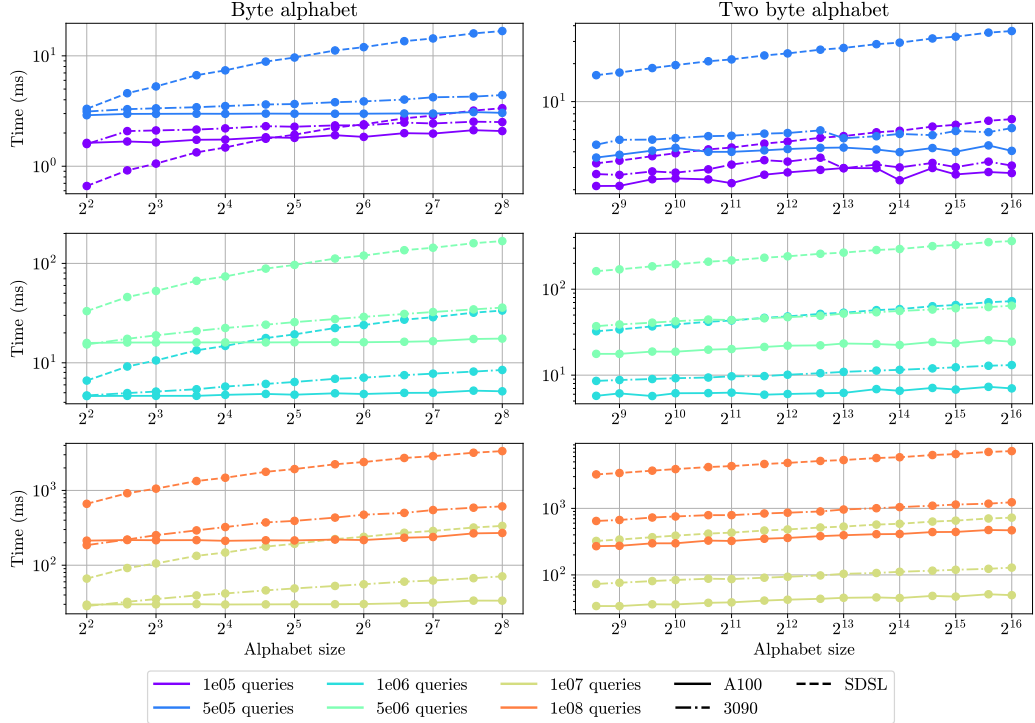}
    \caption{Time taken to process different numbers of access queries coming from the CPU by my implementation and by SDSL, on uniformly randomly distributed texts of 6GB in size for a variety of alphabet sizes.}
    \label{fig:wt_access_random}
\end{figure}

To obtain a more comprehensive view of the performance of the implementations, I will first discuss the results for uniformly randomly distributed texts with a fixed size for a variety of alphabet sizes, shown in Figure \ref{fig:wt_access_random}. It is important to note that, in this case, SDSL was timed only for the 10 million queries case, and the times for other numbers of queries are inferred from those results.

Overall, we can see that on the 3090 and the A100 the time barely increases with increasing alphabet size, which is most probably due to the copying of the queries, which is independent of alphabet size, being the main bottleneck of the operation and not the processing. Interestingly, as the total number of queries increases, the 3090 adopts a more logarithmic shape, while the A100 remains mostly constant. This is probably due to the processing of the queries becoming the main bottleneck of the operation on the 3090. On SDSL, the logarithmic increase in execution time is clearly visible. 

For 100'000 queries, the A100 becomes faster for $\sigma \geq 32$, and the 3090 for $\sigma \geq 64$. From 500'000 queries onwards, both the A100 and the 3090 are always faster than SDSL. For 100 million queries, where the parallelism of the GPUs is fully utilized, the 3090 is 3.5x to 5.9x faster, and the A100 3.1x to 15.5x faster than SDSL, where the difference increases with the alphabet size.

\begin{figure}[H]
    \centering
    \includesvg[width=.9\linewidth]{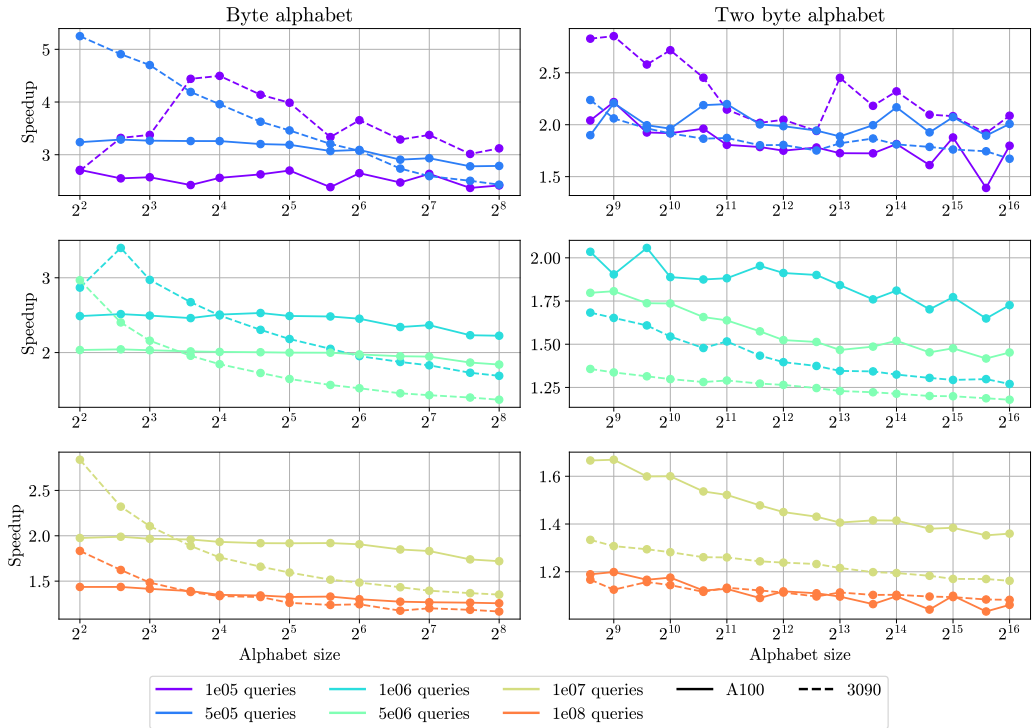}
    \caption{Effect of allocating access queries as pinned memory when processing different numbers of them coming from the CPU, on uniformly randomly distributed texts of 6GB in size for a variety of alphabet sizes.}
    \label{fig:wt_access_random_pinned}
\end{figure}

Figure \ref{fig:wt_access_random_pinned} shows the effects of not having to pin the queries if they have been allocated as pageable memory. Generally, we can see that the achieved speedup decreases as the alphabet size increases. As expected, the achieved speedup also decreases as the number of queries increases since the overhead of having to pin the queries becomes less noticeable. Interestingly, for the byte alphabet, the achieved speedup on the 3090 tends to decrease considerably as the alphabet size increases, while on the A100 it remains mostly constant. 
This may be because as the alphabet size increases, the time to process the queries increases as well, while the time necessary to pin the queries does not. Though the results seem noisy and indicative of inadequate benchmarking, they were collected as the median time from 200 timed iterations, on the exact same wavelet tree and queries, which should minimize the chance of the collected result being an outlier.

\begin{figure}[H]
    \centering
    \includesvg[width=.9\linewidth]{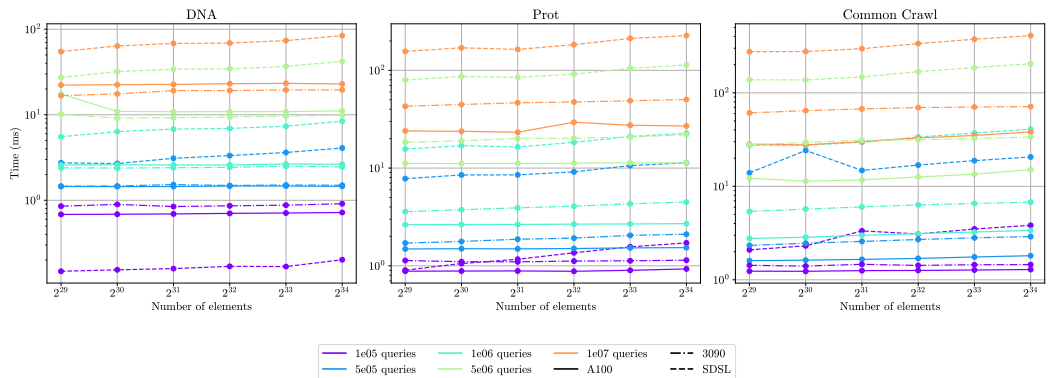}
    \caption{Time taken to process different numbers of access queries coming from the CPU by my implementation and by SDSL, on real-world texts.}
    \label{fig:wt_access_from_data}
\end{figure}

For the real-world texts, the results are shown in Figure \ref{fig:wt_access_from_data}.
On DNA, the GPU im\-ple\-men\-ta\-tion is around five times slower than SDSL for 100'000 queries but becomes around two times faster for 500'000 queries. The difference keeps increasing with the number of queries, and for 10M queries, the GPU implementation is three to five times faster than SDSL, depending on the size of the text. Interestingly, for 100,000 queries, the A100 is faster than the 3090, but as the number of queries increases, the 3090 becomes faster than the A100.

On Prot, already for 100'000 queries, the A100 is faster than SDSL, and the 3090 as well from a text size of $2^{31}$ elements onwards. For 10M queries, the A100 is at least four times faster, and the 3090 at least 2.5x faster than SDSL. In this case, the A100 is always faster than the 3090, the difference being as much as 2x for large numbers of queries.

On Common Crawl, again starting from 100'000 queries, the A100 is at least 2.5x faster, and the 3090 at least 1.7x faster than SDSL. The difference keeps increasing with the number of queries, and for 10M queries, the A100 is 10 times faster, and the 3090 five times faster than SDSL. As with Prot, the A100 is always faster than the 3090, and the difference increases with the number of queries.

Overall, the A100 and the 3090 scale better with the text size than SDSL, with the execution time barely increasing or not at all.

\begin{figure}[H]
    \centering
    \includesvg[width=.9\linewidth]{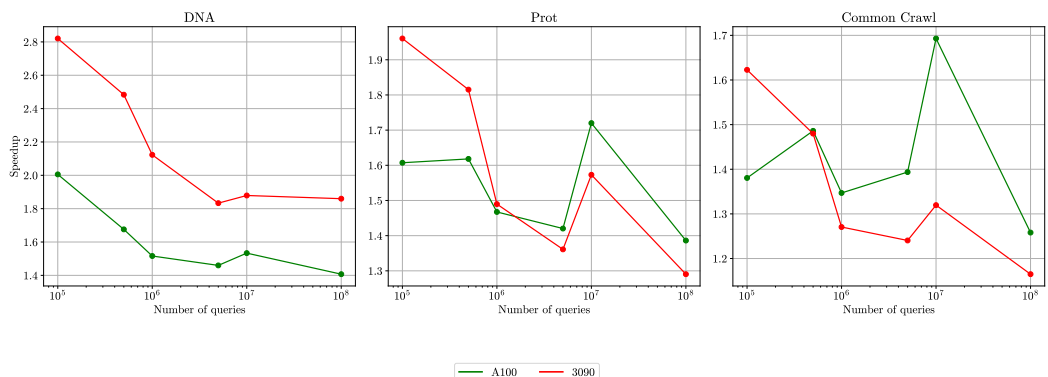}
    \caption{Effect of allocating access queries as pinned memory when processing different numbers of them coming from the CPU, on real-world texts.}
    \label{fig:wt_access_from_data_pinned}
\end{figure}

Figure \ref{fig:wt_access_from_data_pinned} shows the effects of not having to pin the queries if they have been allocated as pageable memory. On DNA, the improvement ranges between 3x and 2x for the 3090 and 2x and 1.4x for the A100. On Prot, it is between 2x and 1.23x for the 3090 and between 1.6x and 1.4x for the A100. On Common Crawl, it is between 1.7x and 1.2x for the 3090 and between 1.4x and 1.25x for the A100, respectively. The improvement decreases as the total number of queries to process increases, since the additional overhead of having to pin the memory becomes less noticeable.

\subsubsection{Rank Query}
\begin{figure}[H]
    \centering
    \includesvg[width=.9\linewidth]{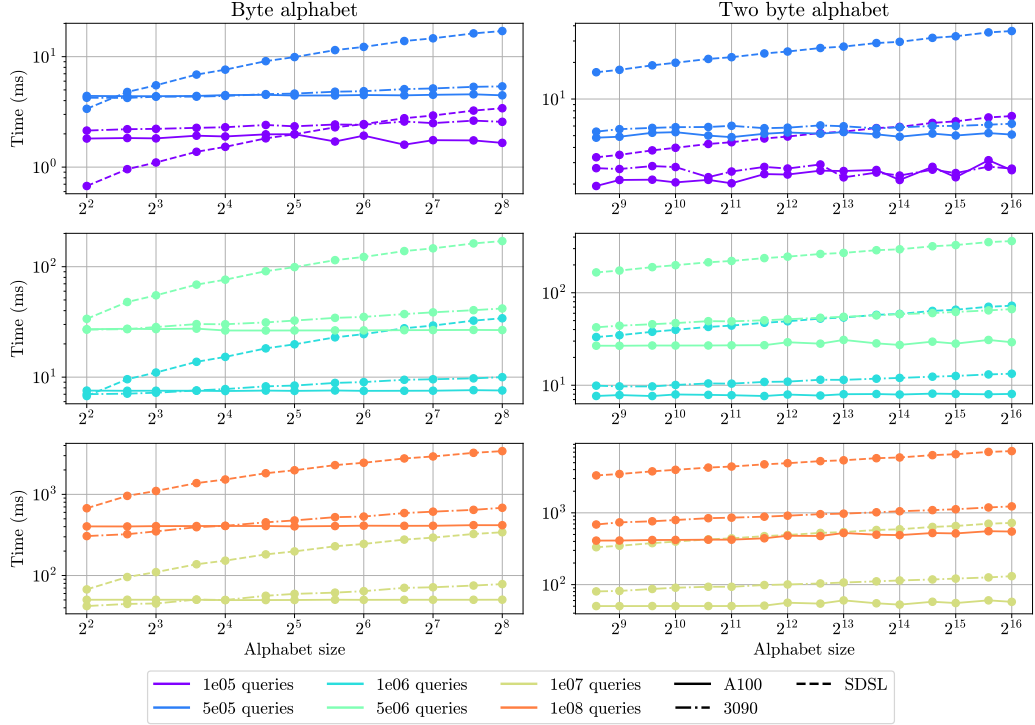}
    \caption{Time taken to process different numbers of rank queries coming from the CPU by my implementation and by SDSL, on uniformly randomly distributed texts of 6GB in size for a variety of alphabet sizes.}
    \label{fig:wt_rank_random}
\end{figure}

As with the access query, we will first look at the results on uniformly randomly distributed texts for a variety of alphabet sizes. These are shown in Figure \ref{fig:wt_rank_random}. Note that although the results for the pinned and sorted queries comparisons that we will see later seem noisy and indicative of inadequate benchmarking, they were collected as the median time from 200 timed iterations, on the exact same wavelet tree and queries, which should minimize the chance that the collected result is an outlier.

Overall, we see the same patterns as with the access query. On the A100 and the 3090 the time barely increases with the alphabet size, except on the 3090 for large numbers of queries, where it does adopt a more logarithmic shape.

For 100'000 queries, the A100 becomes faster for $\sigma \geq 32$, and the 3090 for $\sigma \geq 64$. For 500'000 queries, both the A100 and the 3090 are faster than SDSL for $\sigma \geq 6$. For 100 million queries, where the parallelism of the GPUs is fully utilized, the 3090 is 2.2x to 6x faster, and the A100 1.7x to 13.2x faster than SDSL, the difference increasing with the alphabet size.

\begin{figure}[H]
    \centering
    \includesvg[width=.9\linewidth]{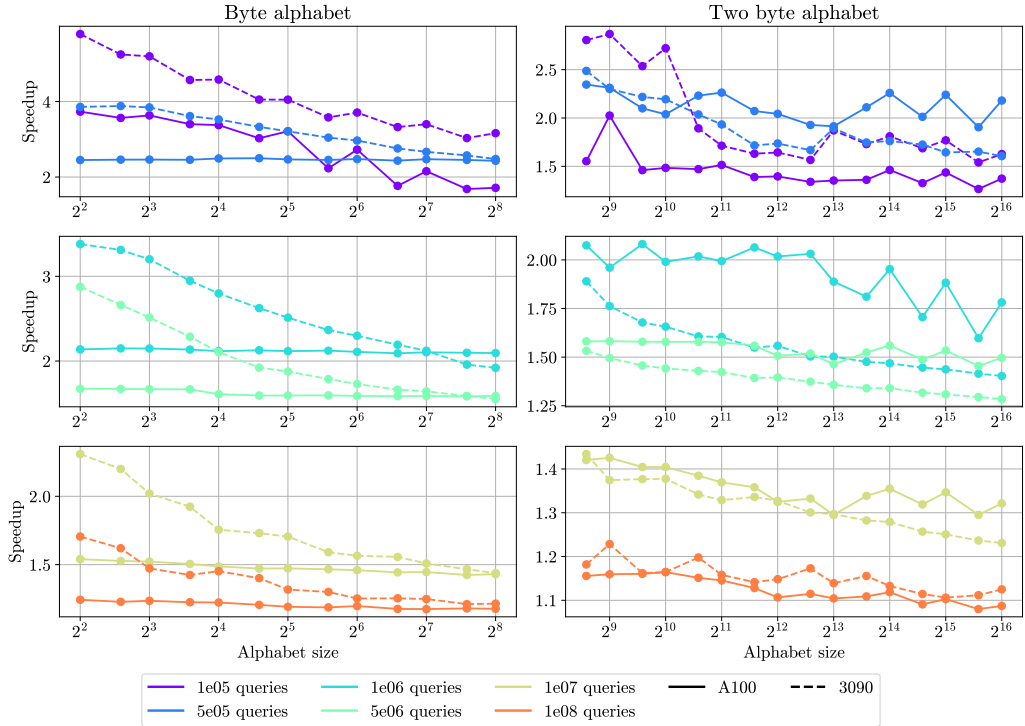}
    \caption{Effect of allocating rank queries as pinned memory when processing different numbers of them coming from the CPU, on uniformly randomly distributed texts of 6GB in size for a variety of alphabet sizes.}
    \label{fig:wt_rank_random_pinned}
\end{figure}

Figure \ref{fig:wt_rank_random_pinned} shows the effects of not having to pin the queries if they have been allocated as pageable memory. They are the same as with the access query, the achieved speedup generally decreases as the alphabet size increases and also as the total number of queries increase.

\begin{figure}[H]
    \centering
    \includesvg[width=.9\linewidth]{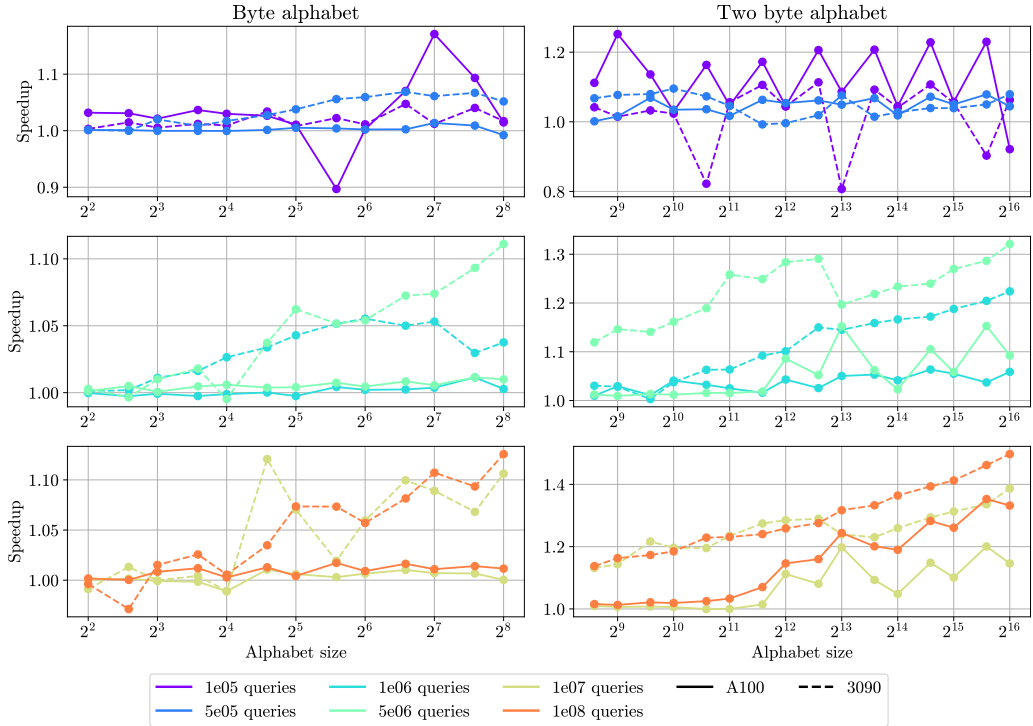}
    \caption{Effect of sorting rank queries by symbol when processing different numbers of them coming from the CPU, on uniformly randomly distributed texts of 6GB in size for a variety of alphabet sizes.}
    \label{fig:wt_rank_random_sorted}
\end{figure}

Something that could also improve the performance of rank queries is sorting them. If we sort the queries by symbol, adjacent threads will often process queries of the same symbol, resulting in them traversing the same nodes of the tree. This can increase cache hit rate and reduce thread divergence. When profiling the query processing kernel by itself, this change improved runtime by 17\% for $\sigma = 150$ and by 35\% for $\sigma = 20'000$. When timing the whole processing function, shown in Figure \ref{fig:wt_rank_random_sorted}, the improvement is not as stark. Note that the sorting time is not included. Generally, the speedup is at most 1.4x, and increases with the alphabet size and with the total number of queries. Interestingly, the speedup tends to be lower on the A100 than on the 3090. This is probably due to the 3090 being more limited by the query processing than the A100, and thus benefitting more from the improved processing throughput. There are also some specific instances where sorting produces a slowdown.

\begin{figure}[H]
    \centering
    \includesvg[width=.9\linewidth]{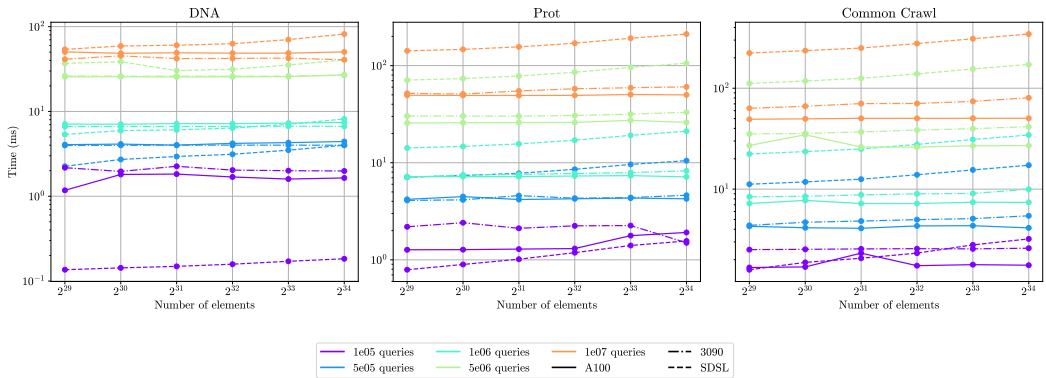}
    \caption{Time taken to process different numbers of rank queries coming from the CPU by my implementation and by SDSL, on real-world texts.}
    \label{fig:wt_rank_from_data}
\end{figure}

For the real-world texts, the results are shown in Figure \ref{fig:wt_rank_from_data}.
On DNA, the GPU im\-ple\-men\-ta\-tion needs a high number of queries in order to become faster than SDSL. It is only consistently faster from 5M queries onwards. For that amount of queries, both the A100 and the 3090 are around 1.4x faster on average. For 10M queries, the A100 is between 1.1x and 1.7x faster, and the 3090 between 1.4x and 2x faster. When comparing the A100 and the 3090, there is the same behavior as with the access query, where the A100 is faster for 100'000 queries, but the 3090 overtakes it as the queries increase.

On Prot, the GPU implementation is faster from 500'000 queries onwards. For this amount of queries, both the A100 and the 3090 are between 1.8x and 2.3x faster. For 10M queries, they are between 2.8x and 3.6x faster. When comparing the A100 and the 3090, the A100 is usually faster, but by a small margin.

On Common Crawl, already for 100'000 queries, the A100 is 1.3x faster on average, and the 3090 becomes faster for a text size $n \geq 2^{33}$. The difference keeps increasing with the number of queries, and for 10M queries, the A100 is 4.5x to 7x faster, and the 3090 3.5x to 4.3x faster than SDSL. The A100 is always faster than the 3090 in this case.

Overall, as with the access query, the A100 and the 3090 scale better with the text size than SDSL, execution time barely increasing, or not at all.

\begin{figure}[H]
    \centering
    \includesvg[width=.9\linewidth]{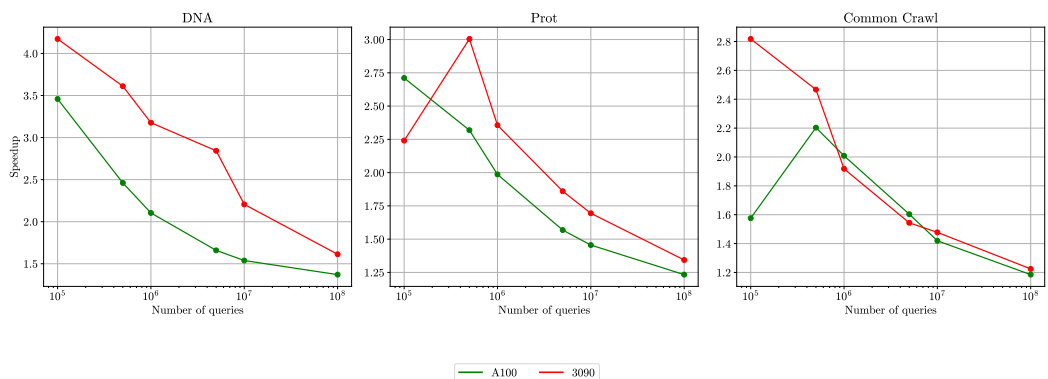}
    \caption{Effect of allocating rank queries as pinned memory when processing different numbers of them coming from the CPU, on real-world texts.}
    \label{fig:wt_rank_from_data_pinned}
\end{figure}

Figure \ref{fig:wt_rank_from_data_pinned} shows the effects of not having to pin the queries if they have been allocated as pageable memory. On DNA, the improvement ranges between 4.5x and 1.6x for the 3090, and between 3.5x and 1.4x for the A100. On Prot, between 2.3x and 1.4x for the 3090, and between 2.7x and 1.3x for the A100. On Common Crawl, between 2.8x and 1.2x for the 3090, and between 1.6x and 1.2x for the A100. As with the access query, the improvement usually decreases as the total number of queries to process increases, since the additional overhead of having to pin the memory becomes less noticeable.

\subsubsection{Select Query}
\begin{figure}[H]
    \centering
    \includesvg[width=.9\linewidth]{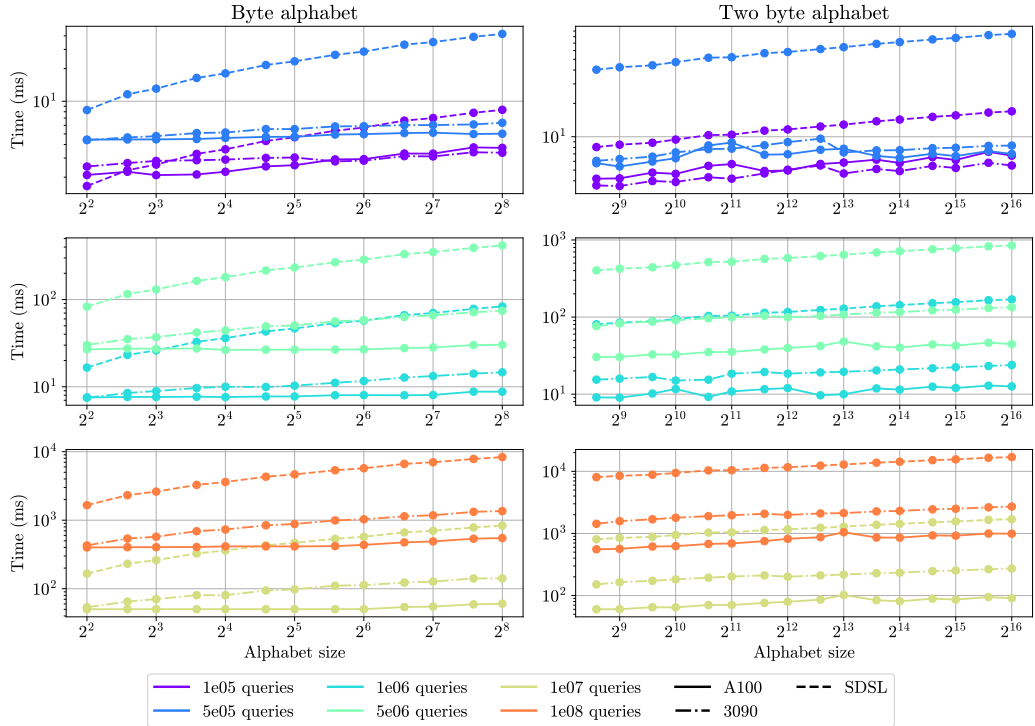}
    \caption{Time taken to process different numbers of select queries coming from the CPU by my implementation and by SDSL, on uniformly randomly distributed texts of 6GB in size for a variety of alphabet sizes.}
    \label{fig:wt_select_random}
\end{figure}

As with the previous queries, we will first look at the results on uniformly randomly distributed texts for a variety of alphabet sizes. These are shown in Figure \ref{fig:wt_select_random}. Note that although the results for the pinned and sorted queries comparisons that we will see later seem noisy and indicative of inadequate benchmarking, they were collected as the median time from 200 timed iterations, on the exact same wavelet tree and queries, which should minimize the chance that the collected result is an outlier.

In contrast to the other queries, here both the 3090's and the A100's time evolution has a more logarithmic shape. This possibly indicates that the copying of the queries is no longer the main bottleneck of the operation, but rather the processing.

For 100'000 queries, the A100 becomes faster for $\sigma \geq 6$, and the 3090 for $\sigma \geq 12$. For 500'000 queries, both the A100 and the 3090 are at least 2x faster than SDSL. For 100 million queries, where the parallelism of the GPUs is fully utilized, the 3090 is 3.9x to 6.3x faster, and the A100 4.2x to 17x faster than SDSL, the difference increasing with the alphabet size.

\begin{figure}[H]
    \centering
    \includesvg[width=.9\linewidth]{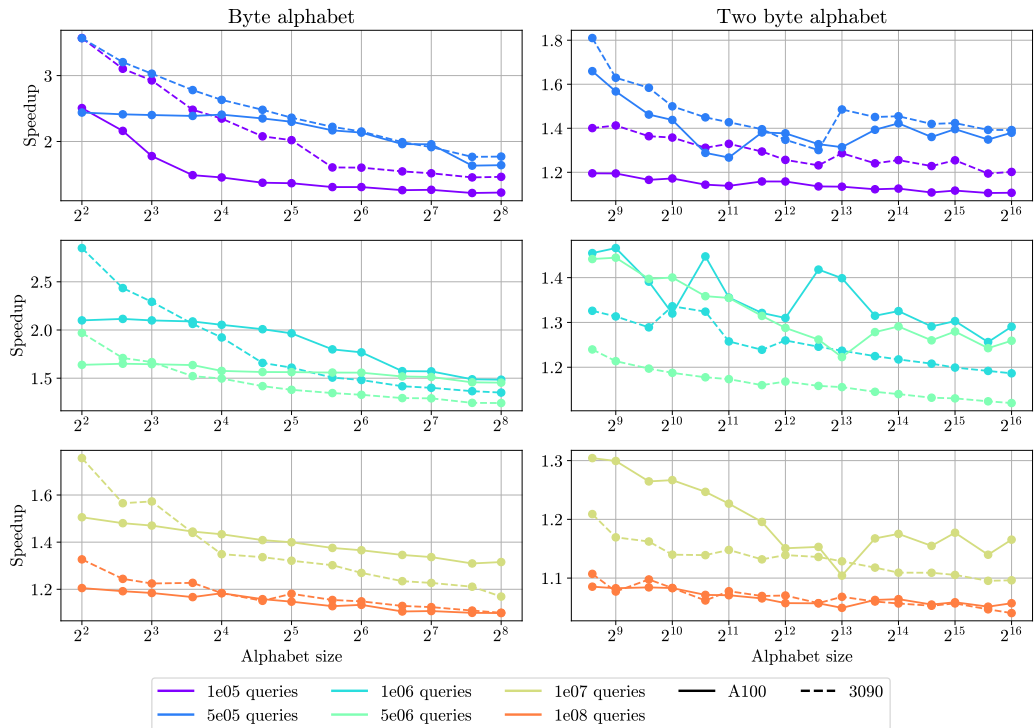}
    \caption{Effect of allocating select queries as pinned memory when processing different numbers of them coming from the CPU, on uniformly randomly distributed texts of 6GB in size for a variety of alphabet sizes.}
    \label{fig:wt_select_random_pinned}
\end{figure}

Figure \ref{fig:wt_select_random_pinned} shows the effects of not having to pin the queries if they have been allocated as pageable memory. They are the same as with the other queries, the achieved speedup decreases as the alphabet size increases and also as the total number of queries increase.

\begin{figure}[H]
    \centering
    \includesvg[width=.9\linewidth]{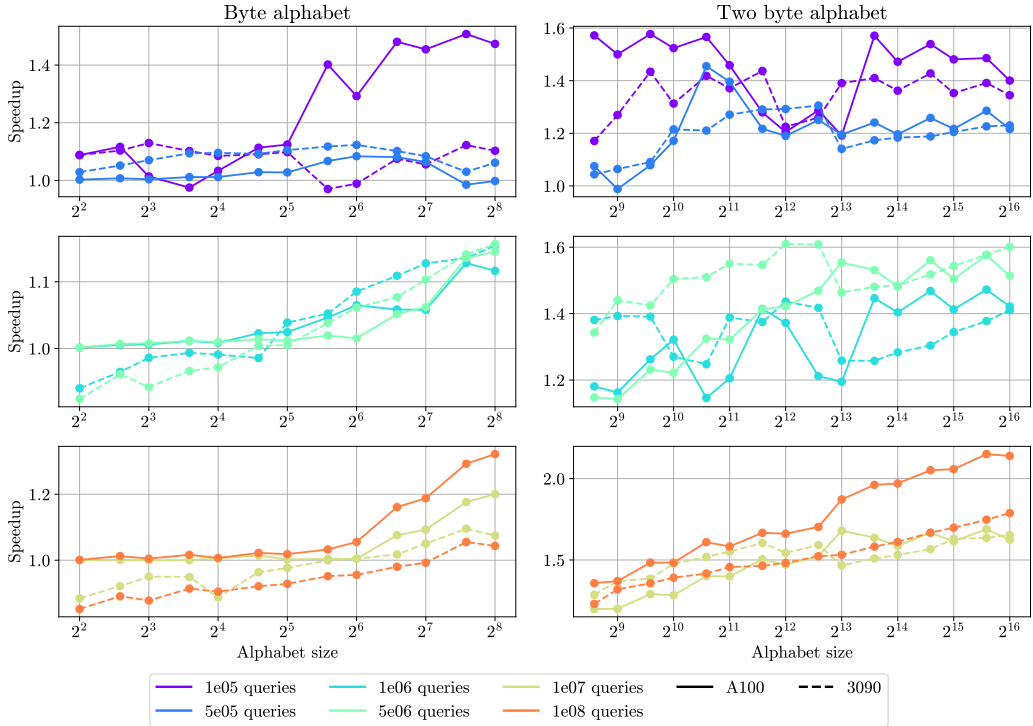}
    \caption{Effect of sorting select queries by symbol when processing different numbers of them coming from the CPU, on uniformly randomly distributed texts of 6GB in size for a variety of alphabet sizes.}
    \label{fig:wt_select_random_sorted}
\end{figure}

Something that could also improve the performance of select queries is sorting them. If we sort the queries by symbol, adjacent threads will process queries of the same symbol, resulting in them traversing the same nodes of the tree, which can increase cache hit rate and reduce thread divergence. When profiling the query processing kernel by itself, this change improved runtime by 18\% for $\sigma = 150$ and by 36\% for $\sigma = 20'000$. When timing the whole processing function, shown in Figure \ref{fig:wt_select_random_sorted}, the improvement is similar. Note that the sorting time is not included. Generally, the speedup is at most 2.2x, and increases with the alphabet size and with the total number of queries. Interestingly, the speedup now tends to be higher on the A100 than on the 3090, and in some cases sorting actually produces a slowdown. A valid hypothesis for this I have not been able to formulate.

\begin{figure}[H]
    \centering
    \includesvg[width=.9\linewidth]{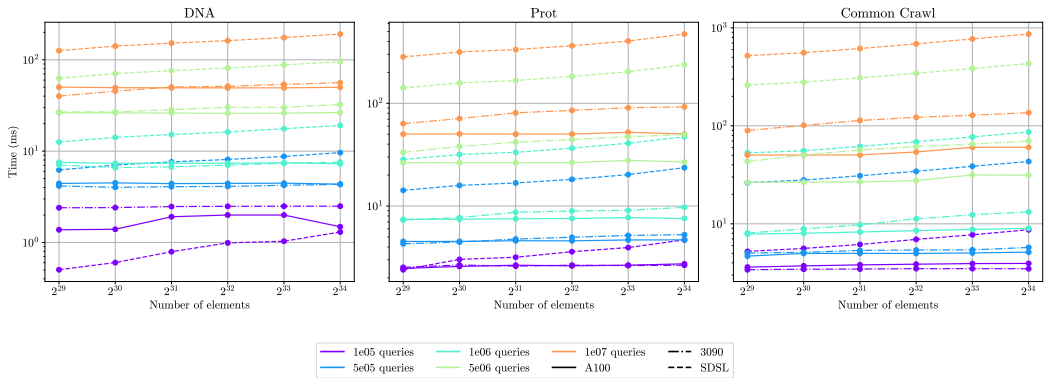}
    \caption{Time taken to process different numbers of select queries coming from the CPU by my implementation and by SDSL, on real-world texts.}
    \label{fig:wt_select_from_data}
\end{figure}

For the real-world texts, the results are shown in Figure \ref{fig:wt_select_from_data}.
On DNA, the GPU im\-ple\-men\-ta\-tion becomes faster than SDSL from 500'000 queries onwards. For that amount of queries, both the A100 and the 3090 are at least 1.4x faster. For 10M queries, the A100 is between 2.5x and 3.8x faster, and the 3090 between 3.1x and 3.4x faster. When comparing the A100 and the 3090, their perfomance is very similar except for the 100'000 queries case, where the A100 around 1.5x faster than the 3090.

On Prot, the GPU implementation is already faster for 100'000 queries. For 10M queries, the A100 is between 5.7x and 10.5x faster, and the 3090 between 4.5x and 5.7x faster. When comparing the A100 and the 3090, the A100 becomes faster as the number of queries increases.

On Common Crawl, already for 100'000 queries, the A100 and 3090 are at least 1.5x faster. The difference keeps increasing with the number of queries, and for 10M queries, the A100 is 10.4x to 14x faster, and the 3090 5.8x to 6x faster than SDSL. For 100'000 queries, the 3090 is slightly faster than the A100, but from then on the A100 becomes faster, up to over 2x faster for large numbers of queries.

Overall, the A100's execution time is largely independent of the text size, while the 3090 starts to show a dependence, especially on Common Crawl. This could be because for the select query, as we saw on Figure \ref{fig:wt_select_random}, the processing seemingly becomes the main bottleneck and not the copying, more so on the 3090. The throughput of the memory copies is not dependent of the size of the text, while that of processing the queries is, as when the text is larger the queries will be more spread out and therefore the cache locality will be reduced.

\begin{figure}[H]
    \centering
    \includesvg[width=.9\linewidth]{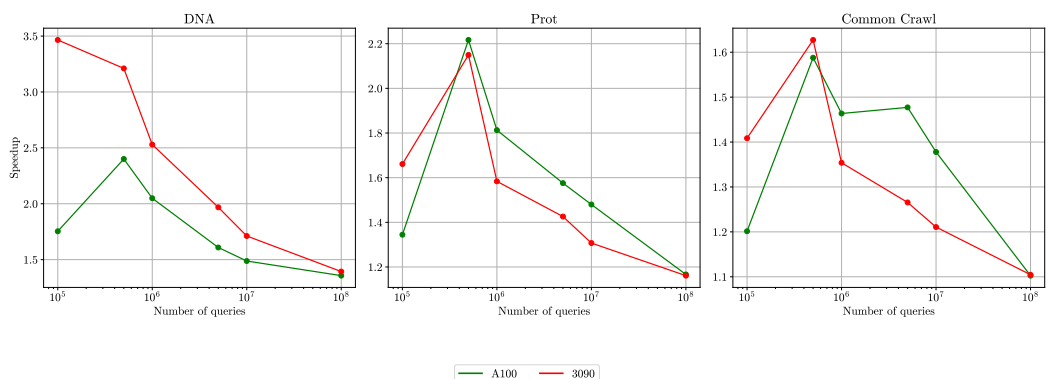}
    \caption{Effect of allocating select queries as pinned memory when processing different numbers of them coming from the CPU, on real-world texts.}
    \label{fig:wt_select_from_data_pinned}
\end{figure}

Figure \ref{fig:wt_select_from_data_pinned} shows the effects of not having to pin the queries if they have been allocated as pageable memory. On DNA, the improvement ranges between 3.5x and 1.4x for the 3090, and between 2.4x and 1.4x for the A100. On Prot, between 2.2x and 1.2x for both. On Common Crawl, between 1.6x and 1.1x for both. As with the other queries, the improvement usually decreases as the total number of queries to process increases, since the additional overhead of having to pin the memory becomes less noticeable.

\newpage
\section{Conclusion and Future Work}
In this thesis, I have presented a new GPU wavelet tree implementation. The goal of the thesis was to achieve an implementation that would outperform the current best publicly available CPU implementations, given enough parallelism, while offering a simple and easily usable API that allows the user of the library to take advantage of the GPU without having to interact with it. 

First, I explained the implementation and optimization process of the binary rank and select support structures, which are the main building blocks of the tree. Compared to their CPU counterpart, these structures allow for at least 10 times higher throughput on binary rank queries (Figure \ref{fig:RS_rank}), and at least 150 times more throughput on binary select queries (Figure \ref{fig:RS_select}), while scaling better with the size of the bit array. For the construction of the structures, I presented the first parallel algorithm, which constructs the structures between 13 and 100 times faster than the best sequential algorithm from prior work (Figure \ref{fig:RS_construction}).

Following that, I explained a new parallel wavelet tree construction algorithm based on the radix sort, and outlined the optimization process. The algorithm itself outperforms the current state of the art with similar construction overhead, but the implementation suffers from the overhead of copying the data the wavelet tree must be built upon from the CPU to the GPU.

Lastly, I describe the implementation and optimization process of the access, rank, and select queries, introducing an optimization to the access algorithm that allows one level less to be traversed. Given enough parallelism, the implementation outperforms the widely used SDSL CPU implementation when passing in the queries from the CPU, allowing a seamless transition from the SDSL library's implementation to this one with minor changes to the code from the side of the user.

Though this wavelet tree implementation achieves good performance, there is still room for improvement. Its biggest weakness is the construction. One improvement that could be made is constructing the tree from the data in chunks, in order to hide the latency of the data copy from the CPU to the GPU. This would also allow us to build the wavelet tree from bigger texts without having to resort to unified memory, which slows down the construction considerably. A custom radix sort implementation would also reduce the additional memory needed for the construction, while possibly improving performance compared to the CUB library.

Apart from improving the tree implementation, the next logical steps would be to implement the wavelet matrix~\cite{wavelet_matrix} and the Huffman versions of the tree and matrix.

\newpage
\bibliographystyle{unsrt}
\bibliography{bibl}

\begin{thebibliography}{10}

\bibitem{wt_for_all}
Gonzalo Navarro.
\newblock Wavelet trees for all.
\newblock {\em Journal of Discrete Algorithms}, 25:2--20, 2014.
\newblock 23rd Annual Symposium on Combinatorial Pattern Matching.

\bibitem{quadWT}
Matteo Ceregini, Florian Kurpicz, and Rossano Venturini.
\newblock Faster wavelet trees with quad vectors.
\newblock Technical report, 2023.

\bibitem{fm-index}
Paolo Ferragina and Giovanni Manzini.
\newblock Indexing compressed text.
\newblock {\em J. ACM}, 52(4):552–581, July 2005.

\bibitem{Bowtie}
Ben Langmead and Steven~L. Salzberg.
\newblock Fast gapped-read alignment with bowtie 2.
\newblock {\em Nature Methods}, 9(4):357--359, Apr 2012.

\bibitem{sequencing}
Scott~D. Kahn.
\newblock On the future of genomic data.
\newblock {\em Science}, 331(6018):728--729, 2011.

\bibitem{Kurpicz_RS}
Florian Kurpicz.
\newblock Engineering compact data structures for rank and select queries on bit vectors.
\newblock In {\em SPIRE}, volume 13617 of {\em Lecture Notes in Computer Science}, page 257–272. Springer, 2022.

\bibitem{SDSL}
Simon Gog, Timo Beller, Alistair Moffat, and Matthias Petri.
\newblock From theory to practice: Plug and play with succinct data structures.
\newblock In Joachim Gudmundsson and Jyrki Katajainen, editors, {\em Experimental Algorithms}, pages 326--337, Cham, 2014. Springer International Publishing.

\bibitem{wavelet_matrix}
Francisco Claude, Gonzalo Navarro, and Alberto Ordóñez.
\newblock The wavelet matrix: An efficient wavelet tree for large alphabets.
\newblock {\em Information Systems}, 47:15--32, 2015.

\bibitem{wt_og}
Roberto Grossi, Ankur Gupta, and Jeffrey~Scott Vitter.
\newblock High-order entropy-compressed text indexes.
\newblock In {\em Proceedings of the Fourteenth Annual ACM-SIAM Symposium on Discrete Algorithms}, SODA '03, page 841–850, USA, 2003. Society for Industrial and Applied Mathematics.

\bibitem{Makris2012-bi}
Christos Makris.
\newblock Wavelet trees: A survey.
\newblock {\em Comput. Sci. Inf. Syst.}, 9(2):585--625, 2012.

\bibitem{practical_WT}
Patrick Dinklage, Jonas Ellert, Johannes Fischer, Florian Kurpicz, and Marvin Löbel.
\newblock Practical wavelet tree construction.
\newblock page 1.8:1–1.8:67, 2021.

\bibitem{WT_compr}
Paolo Ferragina, Raffaele Giancarlo, and Giovanni Manzini.
\newblock The myriad virtues of wavelet trees.
\newblock {\em Information and Computation}, 207(8):849--866, 2009.

\bibitem{Grossi_2003}
Roberto Grossi, Ankur Gupta, and Jeffrey~Scott Vitter.
\newblock High-order entropy-compressed text indexes.
\newblock In {\em Proceedings of the Fourteenth Annual ACM-SIAM Symposium on Discrete Algorithms}, SODA '03, page 841–850, USA, 2003. Society for Industrial and Applied Mathematics.

\bibitem{MAKINEN2007332}
Veli Mäkinen and Gonzalo Navarro.
\newblock Rank and select revisited and extended.
\newblock {\em Theoretical Computer Science}, 387(3):332--347, 2007.
\newblock The Burrows-Wheeler Transform.

\bibitem{CCCL}
{CCCL Development Team}.
\newblock {\em {CCCL}: {CUDA} {C++} {C}ore {L}ibraries}, 2023.

\bibitem{entropy}
S.~Rao Kosaraju and Giovanni Manzini.
\newblock Compression of low entropy strings with lempel--ziv algorithms.
\newblock {\em SIAM Journal on Computing}, 29(3):893--911, 2000.

\end{thebibliography}
\newpage
\appendix
\appendixpagenumbering
\counterwithin{figure}{section}
% An die letzte Seite werden die Nutzungsrechte für studentische Arbeiten vom IFB angehängt.
%\newpage
% \includepdf[pages=-]{Nutzungsrechte_studentische_Arbeit_englisch[6857].pdf}
\end{document}